\documentclass[sigconf]{acmart}
\usepackage[T1]{fontenc}
\usepackage{afterpage}

\usepackage{listings}
\usepackage{xcolor,colortbl}
\usepackage{graphicx}
\usepackage{float}
\usepackage[font=footnotesize]{caption}
\usepackage[font=footnotesize]{subcaption}
\usepackage{placeins}
\usepackage{url}
\usepackage{epstopdf}
\usepackage{booktabs}
\usepackage{footnote}

\makesavenoteenv{table}
\makesavenoteenv{tabular}

\usepackage[acronyms,nonumberlist,nopostdot,nomain,nogroupskip]{glossaries}

\usepackage{tabularx}

\usepackage{paralist}
\usepackage{dblfloatfix}    % To enable figures at the bottom of page

\def \PLOS{
	       \begin{equation}
	       P_{\rm LOS}(d_o)=
       \begin{cases}
       	\min\{1,1.05 e^{-0.0114d_o}\} & \text{if urban}\\
       	\min\{1,(2.1\cdot10^{-6})d_o^2-0.002d_o+1.02\} & \text{if highway}
       	\label{eq:PL_LOS}
       	\end{cases}
       \end{equation} 
}

\newacronym{3gpp}{3GPP}{3rd Generation Partnership Project}
\newacronym{adc}{ADC}{Analog to Digital Converter}
\newacronym{5g}{5G}{5th generation}
\newacronym{aimd}{AIMD}{Additive Increase Multiplicative Decrease}
\newacronym{am}{AM}{Acknowledged Mode}
\newacronym{amc}{AMC}{Adaptive Modulation and Coding}
\newacronym{aqm}{AQM}{Active Queue Management}
\newacronym{awgn}{AGWN}{Additive White Gaussian Noise}
\newacronym{balia}{BALIA}{Balanced Link Adaptation}
\newacronym{bdp}{BDP}{Bandwidth-Delay Product}
\newacronym{bf}{BF}{Beamforming}
\newacronym{cc}{CC}{Congestion Control}
\newacronym{cdf}{CDF}{Cumulative Distribution Function}
\newacronym{cn}{CN}{Core Network}
\newacronym{cqi}{CQI}{Channel Quality Information}
\newacronym{cp}{CP}{Control Plane}
\newacronym{csirs}{CSI-RS}{Channel State Information - Reference Signal}
\newacronym{dc}{DC}{Dual Connectivity}
\newacronym{dce}{DCE}{Direct Code Execution}
\newacronym{dci}{DCI}{Downlink Control Information}
\newacronym{dl}{DL}{Downlink}
\newacronym{dmr}{DMR}{Deadline Miss Ratio}
\newacronym{dmrs}{DMRS}{DeModulation Reference Signal}
\newacronym{bsm}{BSM}{Basic Safety Message}
\newacronym{cam}{CAM}{Cooperative Awareness Message}
\newacronym{e2e}{E2E}{End-to-End}
\newacronym{ecn}{ECN}{Explicit Congestion Notification}
\newacronym{edf}{EDF}{Earliest Deadline First}
\newacronym{enb}{eNB}{evolved Node Base}
\newacronym{epc}{EPC}{Evolved Packet Core}
\newacronym{es}{ES}{Edge Server}
\newacronym{fdma}{FDMA}{Frequency Division Multiple Access}
\newacronym{fdd}{FDD}{Frequency Division Duplexing}
\newacronym[firstplural=Radio Access Technologies (RATs)]{rat}{RAT}{Radio Access Technology}
\newacronym{fs}{FS}{Fast Switching}
\newacronym{ftp}{FTP}{File Transfer Protocol}
\newacronym{gnb}{gNB}{Next Generation Node Base}
\newacronym{harq}{HARQ}{Hybrid Automatic Repeat reQuest}
\newacronym{hetnet}{HetNet}{Heterogeneous Network}
\newacronym{hh}{HH}{Hard Handover}
\newacronym{hol}{HOL}{Head-of-Line}
\newacronym{ia}{IA}{Initial Access}
\newacronym{imt}{IMT}{International Mobile Telecommunication}
\newacronym{iot}{IoT}{Internet of Things}
\newacronym{los}{LOS}{Line of Sight}
\newacronym{lte}{LTE}{Long Term Evolution}
\newacronym{m2m}{M2M}{Machine to Machine}
\newacronym{mac}{MAC}{Medium Access Control}
\newacronym{mc}{MC}{Multi-Connectivity}
\newacronym{mcs}{MCS}{Modulation and Coding Scheme}
\newacronym{mec}{MEC}{Mobile Edge Cloud}
\newacronym{mi}{MI}{Mutual Information}
\newacronym{mimo}{MIMO}{Multiple Input, Multiple Output}
\newacronym{mmwave}{mmWave}{millimeter wave}
\newacronym{mptcp}{MPTCP}{Multipath TCP}
\newacronym{mr}{MR}{Maximum Rate}
\newacronym{mss}{MSS}{Maximum Segment Size}
\newacronym{mtd}{MTD}{Machine-Type Device}
\newacronym{mtu}{MTU}{Maximum Transmission Unit}
\newacronym{nfv}{NFV}{Network Function Virtualization}
\newacronym{nlos}{NLOS}{Non Line of Sight}
\newacronym{nr}{NR}{NR}
\newacronym{ofdm}{OFDM}{Orthogonal Frequency Division Multiplexing}
\newacronym{pdcch}{PDCCH}{Physical Downlonk Control Channel}
\newacronym{pdcp}{PDCP}{Packet Data Convergence Protocol}
\newacronym{pdsch}{PDSCH}{Physical Downlink Shared Channel}
\newacronym{pdu}{PDU}{Packet Data Unit}
\newacronym{pf}{PF}{Proportional Fair}
\newacronym{pgw}{PGW}{Packet Gateway}
\newacronym{phy}{PHY}{Physical}
\newacronym{pbch}{PBCH}{Physical Broadcast Channel}
\newacronym[plural=\gls{mme}s,firstplural=Mobility Management Entities (MMEs)]{mme}{MME}{Mobility Management Entity}
\newacronym{prb}{PRB}{Physical Resource Block}
\newacronym{pss}{PSS}{Primary Synchronization Signal}
\newacronym{pucch}{PUCCH}{Physical Uplink Control Channel}
\newacronym{pusch}{PUSCH}{Physical Uplink Shared Channel}
\newacronym{rach}{RACH}{Random Access Channel}
\newacronym{ran}{RAN}{Radio Access Network}
\newacronym{red}{RED}{Random Early Detection}
\newacronym{rf}{RF}{Radio Frequency}
\newacronym{rlc}{RLC}{Radio Link Control}
\newacronym{rlf}{RLF}{Radio Link Failure}
\newacronym{rrc}{RRC}{Radio Resource Control}
\newacronym{rrm}{RRM}{Radio Resource Management}
\newacronym{rr}{RR}{Round Robin}
\newacronym{rs}{RS}{Remote Server}
\newacronym{rsrp}{RSRP}{Reference Signal Received Power}
\newacronym{rss}{RSS}{Received Signal Strength}
\newacronym{rtt}{RTT}{Round Trip Time}
\newacronym{rw}{RW}{Receive Window}
\newacronym{rx}{RX}{Receiver}
\newacronym{sa}{SA}{standalone}
\newacronym{sack}{SACK}{Selective Acknowledgment}
\newacronym{sap}{SAP}{Service Access Point}
\newacronym{sch}{SCH}{Secondary Cell Handover}
\newacronym{scoot}{SCOOT}{Split Cycle Offset Optimization Technique}
\newacronym{sdma}{SDMA}{Spatial Division Multiple Access}
\newacronym{sinr}{SINR}{Signal to Interference plus Noise Ratio}
\newacronym{sm}{SM}{Saturation Mode}
\newacronym{snr}{SNR}{Signal to Noise Ratio}
\newacronym{son}{SON}{Self-Organizing Network}
\newacronym{ss}{SS}{Synchronization Signal}
\newacronym{srs}{SRS}{Sounding Reference Signal}
\newacronym{sss}{SSS}{Secondary Synchronization Signal}
\newacronym{tb}{TB}{Transport Block}
\newacronym{tcp}{TCP}{Transmission Control Protocol}
\newacronym{tdd}{TDD}{Time Division Duplexing}
\newacronym{tdma}{TDMA}{Time Division Multiple Access}
\newacronym{tfl}{TfL}{Transport for London}
\newacronym{tm}{TM}{Transparent Mode}
\newacronym{trp}{TRP}{Transmitter Receiver Pair}
\newacronym{tti}{TTI}{Transmission Time Interval}
\newacronym{ttt}{TTT}{Time-to-Trigger}
\newacronym{tx}{TX}{Transmitter}
\newacronym{ue}{UE}{User Equipment}
\newacronym{ul}{UL}{Uplink}
\newacronym{uml}{UML}{Unified Modeling Language}
\newacronym{um}{UM}{Unacknowledged Mode}
\newacronym{utc}{UTC}{Urban Traffic Control}
\newacronym{vm}{VM}{Virtual Machine}
\newacronym{rsrq}{RSRQ}{Reference Signal Received Quality}
\newacronym{rssi}{RSSI}{Received Signal Strength Indicator}
\newacronym{crs}{CRS}{Cell Reference Signal}
\newacronym{v2v}{V2V}{Vehicle-to-Vehicle}
\newacronym{v2i}{V2I}{Vehicle-to-Infrastructure}
\newacronym{v2n}{V2N}{Vehicle-to-Network}
\newacronym{v2x}{V2X}{Vehicle-to-Everything}
\newacronym{vn}{VN}{Vehicular Node}
\newacronym{dsrc}{DSRC}{Dedicated Short Range Communication}
\newacronym{ci}{CI}{context information}
\newacronym{voi}{VoI}{value of information}
\newacronym{aoi}{AoI}{age of information}
\newacronym{gps}{GPS}{Global Positioning System}
\newacronym{qos}{QoS}{Quality of Service}
\newacronym{qoe}{QoE}{Quality of Experience}
\newacronym{ml}{ML}{Machine Learning}
\newacronym{ahp}{AHP}{Analytic Hierarchy Process}
\newacronym{lidar}{LIDAR}{Light Detection and Ranging}
\newacronym{c-its}{C-ITS}{Cooperative and Intelligent Transportation System}
\newacronym{vivat}{VIVAT}{Value of Information in VehiculAr neTwork}
\newacronym{cav}{CAV}{Connected and Automated Vehicle}
\newacronym{ldm}{LDM}{Local Dynamic Map}

\usepackage{tikz}
\usepackage{pgfplots}
\pgfplotsset{compat=newest} 
\pgfplotsset{plot coordinates/math parser=false} 
\newlength\fheight
\newlength\fwidth
\usetikzlibrary{plotmarks,patterns,decorations.pathreplacing,backgrounds,calc,arrows,arrows.meta,spy,matrix}
\usepgfplotslibrary{patchplots,groupplots}
\usepackage{tikzscale}
\usepackage{hyperref}

\usepackage{multirow}
\usepackage{tkz-kiviat}
\tikzstyle{startstop} = [rectangle, rounded corners, minimum width=2cm, minimum height=0.5cm,text centered, draw=black]
\tikzstyle{io} = [trapezium, trapezium left angle=70, trapezium right angle=110, minimum width=3cm, minimum height=1cm, text centered, draw=black]
\tikzstyle{process} = [rectangle, minimum width=2cm, minimum height=0.5cm, text centered, draw=black, alignb=center]
\tikzstyle{decision} = [ellipse, minimum width=2cm, minimum height=1cm, text centered, draw=black]
\tikzstyle{arrow} = [thick,<->,>=stealth]
\tikzstyle{line} = [thick,>=stealth]
\tikzstyle{darrow} = [thick,<->,>=stealth,dashed]
\tikzstyle{sarrow} = [thick,->,>=stealth]
\tikzstyle{larrow} = [line width=0.1mm,dashdotted,->,>=stealth]

\makeatletter
\def\grd@save@target#1{%
  \def\grd@target{#1}}
\def\grd@save@start#1{%
  \def\grd@start{#1}}
\tikzset{
  grid with coordinates/.style={
    to path={%
      \pgfextra{%
        \edef\grd@@target{(\tikztotarget)}%
        \tikz@scan@one@point\grd@save@target\grd@@target\relax
        \edef\grd@@start{(\tikztostart)}%
        \tikz@scan@one@point\grd@save@start\grd@@start\relax
        \draw[minor help lines] (\tikztostart) grid (\tikztotarget);
        \draw[major help lines] (\tikztostart) grid (\tikztotarget);
        \grd@start
        \pgfmathsetmacro{\grd@xa}{\the\pgf@x/1cm}
        \pgfmathsetmacro{\grd@ya}{\the\pgf@y/1cm}
        \grd@target
        \pgfmathsetmacro{\grd@xb}{\the\pgf@x/1cm}
        \pgfmathsetmacro{\grd@yb}{\the\pgf@y/1cm}
        \pgfmathsetmacro{\grd@xc}{\grd@xa + \pgfkeysvalueof{/tikz/grid with coordinates/major step x}}
        \pgfmathsetmacro{\grd@yc}{\grd@ya + \pgfkeysvalueof{/tikz/grid with coordinates/major step y}}
        \foreach \x in {\grd@xa,\grd@xc,...,\grd@xb}
        \node[anchor=north] at (\x,\grd@ya) {\pgfmathprintnumber{\x}};
        \foreach \y in {\grd@ya,\grd@yc,...,\grd@yb}
        \node[anchor=east] at (\grd@xa,\y) {\pgfmathprintnumber{\y}};
      }
    }
  },
  minor help lines/.style={
    help lines,
    gray,
    line cap =round,
    xstep=\pgfkeysvalueof{/tikz/grid with coordinates/minor step x},
    ystep=\pgfkeysvalueof{/tikz/grid with coordinates/minor step y}
  },
  major help lines/.style={
    help lines,
    line cap =round,
    line width=\pgfkeysvalueof{/tikz/grid with coordinates/major line width},
    xstep=\pgfkeysvalueof{/tikz/grid with coordinates/major step x},
    ystep=\pgfkeysvalueof{/tikz/grid with coordinates/major step y}
  },
  grid with coordinates/.cd,
  minor step x/.initial=.5,
  minor step y/.initial=.2,
  major step x/.initial=1,
  major step y/.initial=1,
  major line width/.initial=1pt,
}
\makeatother

\usepackage{makecell}
\usepackage{diagbox}
\usepackage{tikz-qtree}
\usetikzlibrary{trees} % this is to allow the fork right path
\newcolumntype{b}{X}
\newcolumntype{s}{>{\hsize=.01\hsize}X}

\usetikzlibrary{arrows}
\usepackage{enumitem}

\begin{document}
% reduce space before and after eq

%\setlength{\abovedisplayskip}{2pt}
%\setlength{\belowdisplayskip}{2pt}
% space after Figure captions
\setlength{\belowcaptionskip}{-0.4cm}
%\setlength{\headsep}{0.05in}
%\setlength{\topskip}{-10mm}

%align the bottom of the pages
\flushbottom
\setlength{\parskip}{0ex plus0.1ex}

\title{A Framework to Assess Value of Information \\ in Future Vehicular Networks}
%\numberofauthors{3} %  in this sample file, there are a *total*

\author{\vspace{-0.33cm}\texorpdfstring{ Marco Giordani$^{\circ }$, Takamasa Higuchi$^{\dagger }$\thanks{$^{\dagger }$The authors are currently with
Toyota North America R\&D, InfoTech Labs \vspace{0.09cm}},  Andrea Zanella$^{\circ }$,  Onur Altintas$^{\dagger}$, Michele Zorzi$^{\circ }$\\
\vspace{0.09cm}
\small $^{\circ }$ Department of Information Engineering, University of Padova, Padova, Italy, e-mail: \{giordani, zanella, zorzi\}@dei.unipd.it}{} \\
\vspace{-0.39cm}
\small $^{\dagger }$ TOYOTA InfoTechnology Center, USA, Inc., e-mail:{\{ta-higuchi,onur\}@us.toyota-itc.com }}
%\setcopyright{none}
%\settopmatter{printacmref=false, printccs=true, printfolios=true}

\copyrightyear{2019}
%\acmY ear{2019}
\setcopyright{acmcopyright}
\acmConference[TOP-Cars'19]{1st ACM MobiHoc Workshop on Technologies, mOdels, and Protocols for Cooperative Connected Cars }{July 2, 2019}{Catania, Italy} 
%\acmBooktitle{1st ACM MobiHoc Workshop on Technologies, mOdels, and Protocols for Cooperative Connected Cars (TOP-Cars'19), July 2, 2019, Catania, Italy} 
\acmPrice{15.00}
\acmDOI{10.1145/3331054.3331551} \acmISBN{978-1-4503-6807-0/19/07}

% \copyrightyear{2019}
% \setcopyright{acmcopyright}
% \acmConference[TOP-Cars 2019]{the 1st ACM Workshop on Technologies, mOdels, and Protocols for Cooperative Connected Cars (TOP-Cars), co-located with ACM MobiHoc}{July 2019}{Catania, Italy}
% \acmISBN{xxxxxxxxxxxx}
% \acmPrice{15.00}
% \acmDOI{xxxxxxxxxxxxxxxxxx}

\pagestyle{empty}

\begin{abstract}
Vehicles are becoming increasingly intelligent and connected, incorporating more and more sensors to support safer and more efficient driving.
The large volume of data generated by such sensors, however, will likely saturate the capacity of vehicular communication technologies, making it challenging to guarantee the required quality of service.
In this perspective, it is essential to assess the \emph{value of information (VoI)} provided by each data source, to prioritize the transmissions that have the greatest importance for the target applications.
In this paper, we propose and evaluate a framework that uses analytic hierarchy multicriteria decision processes to predict VoI based on  space, time, and quality attributes. 
Our results shed light on the impact of the propagation scenario, the sensor resolution, the type of observation, and the communication distance on the value assessment performance. 
In particular, we show that VoI evolves at different rates as a function of the target application's~characteristics.
\end{abstract}

\begin{CCSXML}
<ccs2012>
 <concept>
  <concept_id>10010520.10010553.10010562</concept_id>
  <concept_desc>Computer systems organization~Embedded systems</concept_desc>
  <concept_significance>500</concept_significance>
 </concept>
 <concept>
  <concept_id>10010520.10010575.10010755</concept_id>
  <concept_desc>Computer systems organization~Redundancy</concept_desc>
  <concept_significance>300</concept_significance>
 </concept>
 <concept>
  <concept_id>10010520.10010553.10010554</concept_id>
  <concept_desc>Computer systems organization~Robotics</concept_desc>
  <concept_significance>100</concept_significance>
 </concept>
 <concept>
  <concept_id>10003033.10003083.10003095</concept_id>
  <concept_desc>Networks~Network reliability</concept_desc>
  <concept_significance>100</concept_significance>
 </concept>
</ccs2012>  
\end{CCSXML}

\ccsdesc[500]{Computer systems organization~Embedded systems}
\ccsdesc[300]{Computer systems organization~Redundancy}
\ccsdesc{Computer systems organization~Robotics}
\ccsdesc[100]{Networks~Network reliability}

\keywords{ Vehicular networking (V2X); value of information (VoI); analytic hierarchy process (AHP)}

\maketitle

\section{Introduction}
\label{sec:introduction}

The automotive industry is  evolving towards \glspl{c-its} to support advanced applications ranging from improved safety to infotainment~\cite{shladover2018connected}. 
Intelligent vehicles will be equipped with sophisticated sensors, including radars, cameras and LIDARs, whose data rate requirements (in the order of terabytes per driving hour~\cite{lu2014connected}) may exceed the capabilities of existing  vehicular communication technologies.

In this scenario, new communication radios operating in the \gls{mmwave} bands above 30 GHz have been investigated because of the large bandwidth available at high frequencies~\cite{magazine2016_Heath,MOCAST_2017}.
Mobile edge computing~\cite{mach2017mobile} and vehicle cloudification~\cite{higuchi2017feasibility}~have also emerged as  promising solutions to meet strict delay requirements  leveraging computation resources at the edge of the network and virtual cloud computing facilities formed by vehicles' on-board computers,~respectively.

%Nonetheless, the challenging radio characteristics of \gls{mmwave} frequencies  may render a significant increase in the available computation capacity ineffective to satisfy the boldest \gls{qos} requirements of future automotive applications~\cite{MOCAST_2017}.
In this context, it is fundamental to set a bound on the amount of information that is distributed over bandwidth-constrained communication channels.
A traditional approach is to monitor the \emph{\gls{aoi}}~\cite{kaul2012real}, i.e., the obsolescence of the data, so that vehicles broadcast sensory messages that are not too old. However, the complex dynamics of vehicular networks affect the rate of decay of the information, making it difficult to set a fixed threshold for the AoI to discriminate between useful and obsolete pieces of data. 
Another approach is to discriminate the \emph{\gls{voi}}~\cite{howard1966information,mason2018feasibility} in order to use the (limited) transmission resources in a way that maximizes the utility for the target applications.

The VoI theory has originally been applied to the military context to prioritize the information to be distributed to soldiers in a battlefield environment~\cite{suri2015exploring}, although such strategies do not account for cases where the information sources are not directly under the users' control.
VoI has also been investigated in underwater systems~\cite{boloni2013scheduling} and \gls{iot} applications~\cite{bisdikian2013quality} to decide how much information to transmit through resource-constrained channels. 
However, while \gls{iot} sensors are mostly static and operate in steady conditions, connected cars are expected to move very rapidly, thereby posing new challenges for proper VoI~characterization.

 In this paper, we investigate for the first time the concept of \gls{voi} in vehicular networks. We therefore propose a method to assess \gls{voi} and rank  scheduling  options as a function of the characteristics of the network in which the nodes are deployed. 
 Traditionally, \gls{voi} assessment strategies, even though not in the vehicular context, try  to balance transmissions of non-critical information and timely dissemination of high-priority data. 
 Some others, e.g.~\cite{bengio2013representation}, use machine learning to assign VoI based on  the degree of correlation among the information sources, although operations are typically computationally heavy, and they can hardly be completed under low latency constraints.
 Adaptive approaches, e.g.~\cite{kamar2013light}, use feedback to predict \gls{voi}, which however incurs non-negligible overhead and communication delays.
Along these lines, a more natural (and practical) solution to assign VoI is to define automotive-specific criteria (i.e., attributes) which depend on the receiver's context and application. 
Our innovative framework, in particular, exploits analytic hierarchy  decision processes (AHP)  to quantify the expected value of information based on time, space and quality interdependencies. 
We validate the technical accuracy of this approach in realistic scenarios and show how  \gls{voi} evolves as a function of the scenario (i.e., urban or highway), the sensor resolution, the type of observation, the communication distance and the age of information.
Thanks to its generality and computational simplicity, our method guarantees timely and efficient value assessment operations.
Through our investigation, we also provide guidelines on the optimal network configurations and data scheduling alternatives that intelligently select information to be sent at each transmission opportunity while maximizing the utility to potential receivers.

 \begin{figure*}[t!]
     \centering
          \centering
     \setlength{\belowcaptionskip}{-0.33cm}
\includegraphics[width=0.95\textwidth]{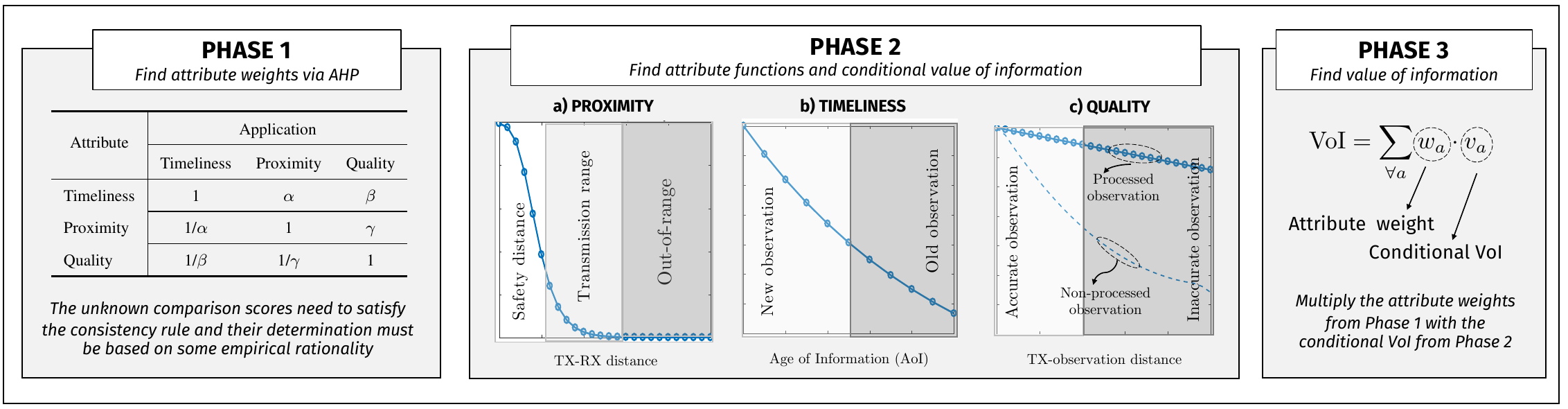}
\vspace{-0.33cm}
\caption{ Illustrative scheme of  VoI assessment framework proposed in Sec.~\ref{sec:voi_assessment_framework}.}
\label{fig:scheme}
\end{figure*}

\vspace{-0.33cm}
\section{VoI in Vehicular Networks} % (fold)
\label{sub:system_model}
VoI~\hspace{0.005cm}is extremely context-dependent, i.e.,~\hspace{0.005cm}different outcomes are~\hspace{0.005cm}possible depending on the~\hspace{0.005cm}characteristics (and  requirements)~\hspace{0.01cm}of~\hspace{0.01cm}the~\hspace{0.01cm}application and~\hspace{0.01cm}the sources of~\hspace{0.005cm}information that are~\hspace{0.005cm}being~considered.

Let $\mathcal{V}$ denote the set of vehicles in the scenario which are equipped with  on-board sensors (e.g., cameras) and a hardware unit enabling \gls{v2v} communications. 
At time  $t$, a vehicle $v_i\in\mathcal{V}$ perceives a road object through its on-board sensors, and generates a perception record $o_{i}(t)$, i.e., the vector of sensors' measurements, for each of the detected objects.
Our key research question in this work is whether vehicle $v_i\in\mathcal{V}$ should broadcast perception record $o_{i}(t)$ to potential receivers.
For example, $o_{i}(t)$ may not be so valuable if destination nodes are not able to receive the perception message in a timely~way. 
We assess that vehicles should distribute the information about each perceived observation only if its  value is above a pre-defined threshold for at least one vehicle within the communication range.
Such a threshold, in turn, may depend on the target application.

%\vspace{-0.33cm}
%\subsection{Vehicular Applications}
 In this paper, we focus on two broad application domains that, for their generality and complementarity, we believe are good representatives of  future vehicular~services.
\begin{itemize}[leftmargin=*]
	\item \emph{Advanced Safety:} It enables semi- or fully-automated driving through dissemination of sensor data, thereby promoting safer traveling and collision avoidance. These applications are typically characterized by  strict latency requirements to  guarantee real-time operations.

\item \emph{Traffic Management:} It enables traffic control and coordination through the creation of a \gls{ldm}~\cite{Eiter2019}, which integrates  sensor data streamed by vehicles in a geographical area.
These operations may require relatively high data rate compared to safety applications, while some latency can be tolerated (depending on the degree of automation).
\end{itemize}%}

In the following, we discuss the information sources and the attribute categories  that affect  \gls{voi}.

\vspace{-0.33cm}
\subsection{Information Sources}
\label{sub:information_sources}
As \glspl{c-its} evolve towards the support of safety-critical applications, it is fundamental to implement network architectures that guarantee timely and accurate positioning of vehicles.
Positioning is typically provided by the \gls{gps} (which also guarantees accurate time synchronization among vehicles), although other localization techniques, e.g., based on image processing, can be useful to improve accuracy.
In this paper, we consider camera sensors as the principal information source to enable position estimation.
The accuracy of the camera observations depends on (i) the resolution of the sensor, which is a measure of the image width/height and the frame rate, (ii) the field of view, and (iii) the operating~distance.
%Notice also that camera observations may be exchanged through cooperative networking among connected cars, even though this may incur in non-negligible delays.
%The data rate demands are proportional to the resolution of the exchanged sensory data and likely exceed 1000 Mbps for high-quality uncompressed camera measurements (e.g., ProRes 4444 with 4K resolution requires around 1200 Mbps). 

%\smallskip
\subsection{Vehicular Attributes}
\label{ssec:attr}
%Considering the transient nature of the vehicular topology, 
\gls{voi} typically decays over time at a rate that is application dependent. Value determination should indeed account for specific attributes, e.g., timeliness, spatial proximity and~quality, as defined next.
\begin{itemize}[leftmargin=*]
\item \emph{Timeliness:} % (fold)
VoI decreases with the relative age of information, i.e., the time between the generation and reception of the information, normalized to its lifetime, i.e., the temporal horizon over which that piece of information is considered valuable. 

\item \emph{Proximity:}
VoI is a function of (i) the distance between the  information source and destination, i.e., sensory data generated by close-by vehicles are generally more valuable than data coming from farther nodes, and (ii) the scenario, i.e., urban/highway.

\item \emph{Information Quality:}
VoI depends on (i) the intrinsic  quality of the information source, which may be assessed~in terms of sensor resolution, and (ii) the distance~between the source sensor and the  observation (e.g., the  depth measurement error increases proportionally with the~distance).
\end{itemize}
% paragraph timeliness (end)

In general,  \gls{voi} can be also affected by network parameters. We do not preclude in the future to further extend the list of attributes included in this paragraph with other factors including, e.g., the vehicle density, the role of a vehicle in the network (e.g., platoon head, cluster coordinator, etc.), and the road topology.

\vspace{-0.33cm}
\section{VoI Assessment Framework} % (fold)
\label{sec:voi_assessment_framework}

In this paper, we consider two possible embodiments of VoI assessment: processed and non-processed approaches. 
The trade off involves latency, energy consumption and VoI accuracy.
In the first case, the perception record is analyzed by the sender to extract context information (e.g., estimated positions of objects in a captured image) before being broadcast. While incurring some processing delays, this allows the sender to validate the integrity of the observation and determine whether it embeds valuable characteristics for the potential receiver(s).
In the second case,  the perception record is broadcast immediately after it is generated, thereby yielding a more responsive value assessment operation. The sender, however, needs to predict  \emph{probabilistically} the value of the observation to prevent  circulation of duplicate or redundant~data.

The proposed framework performs VoI assessment operations through three main phases, as illustrated in Fig.~\ref{fig:scheme} and described~next.

\vspace{-0.33cm}
\subsection*{Phase 1: Attribute Priority Weights (via AHP)} % (fold)
\label{sub:phase_1_ahp_attribute_weights}
First, the algorithm applies the \gls{ahp} to derive the relative degree of priority among the~vehicular attributes, i.e., timeliness, proximity and quality, by  populating a pairwise comparison matrix $M$ (as illustrated in the left frame of Fig.~\ref{fig:scheme}) with comparison scores (i.e., $\alpha$, $\beta$, $\gamma$)~\cite{mu2018understanding}.
The comparison scores~in $M$ (ranging from 1/9 to 9) are  assigned according to the Saaty comparison~scale~\cite{saaty1990decision} and assess the importance of the attributes in the row\hspace{0.06cm}relative\hspace{0.06cm}to\hspace{0.06cm}those in\hspace{0.06cm}the\hspace{0.06cm}column (e.g., the score 3 is  assigned if the item on the row is ``moderately more important'' than the item on the column in the specified application domain).
Note that $M$ is\hspace{0.06cm}reciprocal\hspace{0.06cm}by\hspace{0.06cm}construction, i.e., $M(j,k)=1/M(k,j)$, $\forall j,k\hspace{-0.06cm}\in\hspace{-0.06cm}\{1,\dots,n\}$, where $n$ is\hspace{0.06cm}the\hspace{0.06cm}size\hspace{0.06cm}of\hspace{0.05cm}$M$, i.e., the number of~attributes.

As soon as the comparison scores have been determined, priority weights $w_a$, $a=1,\dots,n$, are computed  evaluating the normalized principal eigenvector $\vec{w}=\langle w_1,\dots,w_n \rangle$ of $M$, i.e., the eigenvector that corresponds to the eigenvalue $\lambda_{\rm max}$ with the largest~magnitude:
\begin{equation}\label{eq:eigenvector}
	M\vec{w}=\lambda_{\rm max}\vec{w}.
\end{equation}
The priority weights indicate  how valuable each attribute is compared to the others.
It appears clear that the AHP method determines relative (instead of absolute) priority weights, which are based on empirical evaluation criteria, and consequently have a certain degree of arbitrariness.
According to the AHP, in order for the weight vector $\vec{w}$ to be a good representative of the relative importance of the attributes, the matrix $M$ should be consistent, i.e., such that $M(j,k)=M(h,k)/M(h,j)$,  $\forall h,j,k \in \{1,\ldots,n\}$. However, given the arbitrariness in the attribute selection, the matrix $M$ is usually not consistent. A measure of the matrix consistency is given in~\cite{mu2018understanding} in terms of the so-called \emph{consistency index} $C_I= (\lambda_{\max}-n)/(n-1)$. Based on~\cite{mu2018understanding}, the weight vector $\vec{w}$ can be considered acceptable if $M$ satisfies the following consistency~rule: 
\begin{equation}\label{eq:consistency}
	C_R = \frac{C_I}{R_I(n)}=\frac{(\lambda_{\rm max}-n)/(n-1)}{R_I(n)} < 0.1,
\end{equation}
where $R_I$ is the average of the $C_I$s obtained by randomly generating reciprocal matrices of size $n$ (for $n=3$, we get $R_I=0.58$).

\subsection*{Phase 2: Conditional VoI} % (fold)
\label{sub:phase_2_conditional_voi}
The framework is now used to assess how  \gls{voi} evolves, conditioned to each attribute. Considering the $n=3$ attributes presented in Sec.~\ref{ssec:attr}, different \gls{voi} functions are~defined.

\begin{figure*}[b!]
     \centering
          \centering
          %\vspace{-1cm}
     \setlength{\belowcaptionskip}{-0.5cm}
\includegraphics[width=0.95\textwidth]{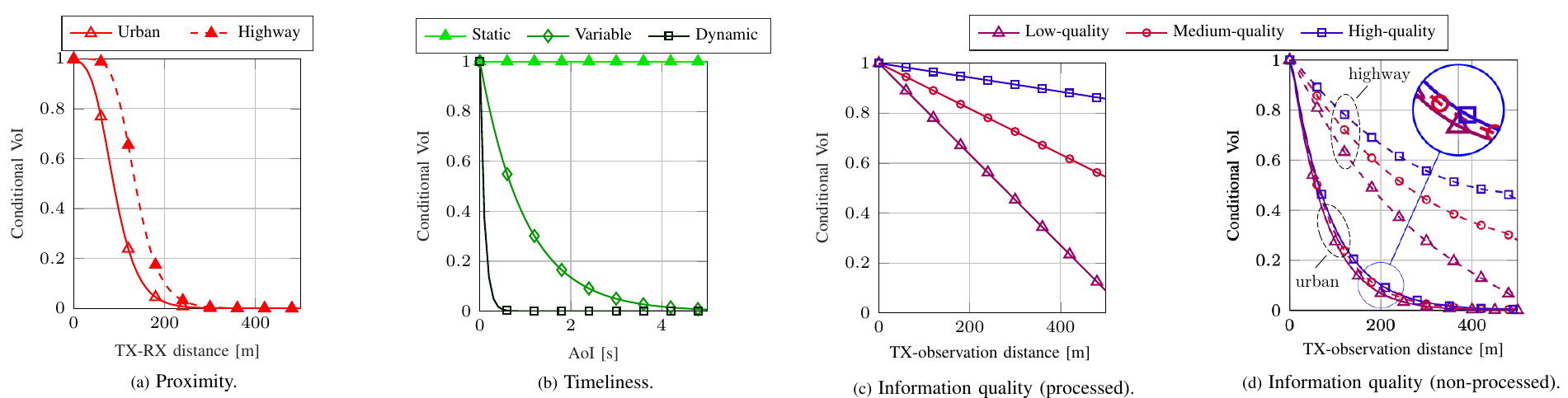}
  %                    \begin{subfigure}[t!]{0.2\textwidth}
  %                    \centering
  %         \setlength{\belowcaptionskip}{0cm}
  % \setlength{\belowcaptionskip}{0cm}
  %   \setlength\fwidth{0.9\columnwidth}
  % \setlength\fheight{0.99\columnwidth}
  % \input{voi_space.tex}
  %   \caption{\footnotesize  Proximity.}
  %   \label{fig:voi_space}
  %     \end{subfigure} \, \qquad \qquad \qquad
  %                  \begin{subfigure}[t!]{0.2\textwidth}
  %                  \centering
  %               \setlength{\belowcaptionskip}{0cm}
  % \setlength{\belowcaptionskip}{0cm}
  %   \setlength\fwidth{0.9\columnwidth}
  % \setlength\fheight{0.99\columnwidth}
  % \input{voi_time.tex}
  %   \caption{\footnotesize Timeliness.}
  %   \label{fig:voi_time}
  %     \end{subfigure}\,
  %      \begin{subfigure}[t!]{0.2\textwidth}
  %                  \centering
  %               \setlength{\belowcaptionskip}{0cm}
  % \setlength{\belowcaptionskip}{0cm}
  %   \setlength\fwidth{0.9\columnwidth}
  % \setlength\fheight{0.99\columnwidth}
  % \input{voi_quality_unprocessed.tex}
  %   \caption{\footnotesize Information quality (processed).}
  %   \label{fig:voi_quality}
  %     \end{subfigure}\,\quad \quad \quad
  %            \begin{subfigure}[t!]{0.2\textwidth}
  %                  \centering
  %               \setlength{\belowcaptionskip}{0cm}
  % \setlength{\belowcaptionskip}{0cm}
  %   \setlength\fwidth{0.9\columnwidth}
  % \setlength\fheight{0.99\columnwidth}
  % \input{voi_quality_processed.tex}
  %   \caption{\footnotesize Information quality (non-processed).}
  %   \label{fig:voi_quality}
  %     \end{subfigure}
  \vspace{-0.33cm}
\caption{Conditional VoI for space proximity, timeliness and information quality attributes. The plots show the impact of the TX-RX distance, the AoI, the TX-observation distance, the propagation scenario, the type of observation and the type of sensor.}
\label{fig:voi_attributes}
\end{figure*}

\smallskip
\emph{a) Proximity:} 
For the proximity attribute, we propose to use a logistic function to model  VoI dependency on the distance $d$ (in meters) between the information source and the destination, as represented in the first box of the middle frame of Fig.~\ref{fig:scheme}. The logistic function is always monotonically decreasing in $d$, but tuning the function's parameters it is possible to move from a smooth and quasi-linear decrease of  VoI with $d$ to a step-like behavior, where  VoI is almost constant within a certain range from the source and suddenly drops to zero beyond that range. The function is given~by: 
\begin{equation} \label{eq:proximity}
v_{\rm 1} = A + \frac{K-A} {\left( C + Q e^{ -B (d-d_s)} \right) ^{1 / \nu }}.
\end{equation}
Parameters in Eq.~\eqref{eq:proximity} are selected so that  $v_1 \simeq 1$ as $d\rightarrow 0$ and $v_1 \simeq 0$ as $d\rightarrow \infty$, i.e., beyond the communication range, and their values will be detailed in Table~\ref{tab:params}.
In particular, 
%$A$ is the lower~asymptote, $K$ is the upper asymptote, $B$ is the growth rate and 
$d_s$ is the safety distance (in meters) and determines the threshold beyond which $v_1$ starts decreasing. $d_s$ is therefore a function of the scenario, i.e., urban/highway, in which the vehicles are deployed.

\smallskip
\emph{b) Timeliness:} % (fold)
For the timeliness attribute, we propose to use an exponential function~\cite{boloni2013scheduling},  as represented in the second box of the middle frame of Fig.~\ref{fig:scheme},~i.e.,
\begin{equation}\label{eq:timeliness}
v_{\rm 2} = \exp(-P_{\rm td}\cdot(t-t_0)),
\end{equation}
where $(t-t_0)$ represents the \gls{aoi} (in seconds), and $P_{\rm td}$ is the temporal decay parameter.
In particular, $P_{\rm td}=0$ indicates that the information is not delay sensitive, while a large $P_{\rm td}$ models a quick obsolescence of the~information.

\smallskip
\emph{c) Quality:}
We propose that  \gls{voi} for the quality attribute evolves  as a function of the distance $d_o$ (in meters) between the source sensor and the perceived observation,  as represented in the third  box of the middle frame of Fig.~\ref{fig:scheme}.
We distinguish between processed ($p$)  and  non-processed ($np$)  VoI assessment operations.
In the processed case,  VoI depends only on the characteristics of the sensor~\cite{stein2003vision}. With reference to the camera sensor, we get 
\begin{equation}
v^{p}_{\rm 3}= 1-\frac{d_o}{h\cdot f_d},
\end{equation}
where $h$ is the height of the sensor, and  $f_d$ is the focal distance.
In particular, $f_d$ depends on the  camera image resolution in the horizontal domain ($r_h$) and the field of view ($f_w$) in degrees, so~that
\begin{equation}\label{eq:fd}
f_d=\frac{r_h/2}{\tan(f_w/180\cdot \pi/2)}.
\end{equation}
In the non-processed case, the sender does not know the content of the perception record, therefore it has to identify a method to probabilistically predict whether such record embeds valuable information. 
We assume that a certain observation (e.g., object) can be detected if it is in \gls{los} with respect to the sensor's field of view.
The \gls{los} probability is a function of $d_o$ and the propagation scenario, and is modeled as in~\cite{37885}:
\medmuskip=-1mu
%\thinmuskip=0mu
\thickmuskip=-1mu
\PLOS
\medmuskip=6mu
%\thinmuskip=0mu
\thickmuskip=6mu
The conditional VoI is finally computed~as
\begin{equation}
v^{np}_{\rm 3} = \left(1-\frac{d_o}{h\cdot f_d}\right)\cdot P_{\rm LOS}(d_o).
\end{equation}

\subsection*{Phase 3: Overall VoI} % (fold)
\label{sub:overall_voi}
Finally, the framework assigns the value of information $v$ by multiplying the attribute weights $w_a$, $a=1,\dots,n$, from Phase 1 with the conditional VoI $v_a$, $a=1,\dots,n$, from Phase 2:
\begin{equation}
	v(d,t,d_o)=\sum_{a=1}^n w_a\cdot v_a.
\end{equation}
A \emph{data scheduler} finally sorts the information products in descending order of values and sequentially forwards them to the surrounding receivers.
The scheduler may also cancel transmissions of information whose value is below a pre-defined threshold~$\theta_v$.

\section{Framework Validation and  Results} % (fold)
\label{sec:framework_validation_and_simulation_results}
In this section we validate the technical soundness of our proposed framework in target use cases. %and shed light on the dynamics and parameters that affect the value assessment operations.
Our results can be used as a basis for evaluating the optimal data scheduling strategy that maximizes the utility of the transmitted information for the final receiver(s).

\begin{table}[t!]
%Different examples of reliable and consistent comparison score triplets ($\alpha,\beta,\gamma$) are considered, to model different interdependency characteristics among the considered attribute categories.}
%\vspace{-0.33cm}
\footnotesize
\centering
\begin{tabular}{ll}
\Xhline{2\arrayrulewidth}
Parameter & Value \\ \Xhline{2\arrayrulewidth}
Sensor height $h$ & 1.2 m  \\ 
Camera Field of view $f_w$ & 70 deg  \\ 
 TX/obs distance $d_o$ & $d/2$  \\
 TX/RX distance $d$ & \{1, \dots, 500\} m \\
 Age of information $t-t_0$ & \{0, \dots, 5\} s  \\
  Safety distance $d_s$ \{urban, highway\}  & \{24,72\} m \\
    Temporal decay $P_{\rm td}$ \{static, variable, dynamic\} & \{0,1,10\}\\
Log. function params \{A,K,C,Q,B,V\} & \{1,\,0,\,1,\,1,\,0.03,\,0.2\}\\
 Camera resolution $r_h$ \{low, medium, high\} & \{640,\,1280,\,4096\} px\\
\Xhline{2\arrayrulewidth}
\end{tabular}
%\caption{ Parametric comparison matrix $M_a$ of \gls{voi} attributes for safety applications. The time-dependency attribute  is assumed to be at least ``moderately more important'' than the sensor's information quality (i.e.,  $\beta>3$, according to the Saaty scale \cite{saaty1990decision}).}
\vspace*{0.5cm}
\caption{System parameters.} 
\label{tab:params}
\vspace*{-0.8cm}
\end{table}

%\subsection{System Parameters} % (fold)
%\label{sub:system_parameters}
The system parameters are based on realistic design considerations and are summarized in Table~\ref{tab:params}.
 For the proximity attribute, we calculate the safety distance $d_s$ between vehicles as $d_s=2\cdot v_{\max}$~\cite{ny2011driving}, where $v_{\max}$ is the speed limit, which we set to 12 m/s and 36 m/s in urban and highway scenarios, respectively. 
%Moreover, we assign to the logistic function parameters in Eq.~\eqref{eq:proximity} the following values: $A = 1$ and $K = 0$ (so that $v_1  \rightarrow 1$ as $d \rightarrow -\infty$ and $v_1  \rightarrow 0$ as $d \rightarrow \infty$), $C = 1$, $Q = 1$, $B = 0.03$, and $\nu = 0.2$.
We also let the distance $d$ between the sender and the receiver vary from 1 to 500~m.
For the timeliness attribute, we consider static (e.g., fixed road construction), variable (e.g., temporary social/political events), and dynamic (e.g., pedestrian crossing the street) observations, which correspond to $P_{\rm td}=0,\,1,\,10$, respectively. We  let the AoI parameter $t-t_0$ vary from 0 to 5 s.
For the quality attribute, we consider low-, medium- and high-quality sensors, which are modeled as $640\times480$, $1280\times720$, and $4096\times780$ pixel cameras, respectively.
The sensor is placed at a distance $h=1.2$ m from the road surface, and the field of view is set to $f_w=70$ degrees. 
We also assume that the target observation is placed at distance $d_o=d/2$ from the camera sensor.
We recall that, in order to exemplify the approach, in this work we focus on the evaluation of  VoI for the position data provided by cameras.

%\subsection{Simulation Results} % (fold)
%\label{sub:simulation_results}
%In this section we validate our VoI framework through simulations.
%Our results can be used as a basis for evaluating the optimal data scheduling strategy that maximizes utility of broadcasted information for final receiver(s).

\smallskip
\textbf{Phase 1 results -- \emph{attribute weights}.}
In Table~\ref{tab:attributes} we report the pairwise comparison matrices $M$ which assess the interdependencies among the considered VoI attributes.\footnote{Notice that the comparative scores we consider in Table~\ref{tab:attributes} are chosen in such a way that the consistency rule defined in Eq.~\eqref{eq:consistency} is~satisfied. 
Other combinations of scores can be also considered, as long as they are selected in a way that guarantees that the assigned attribute interdependencies are fully representative of the characteristics of the application under consideration.}
For instance, for safety applications, we chose to set the proximity vs. timeliness score to $7$ since we deem extremely important for vehicles to monitor space while broadcasting context information (e.g., vehicles need to know when the neighbors' distance falls below the safety-critical threshold to trigger collision avoidance transmissions and, at the same time, should defer or cancel transmissions relative to spatially far vehicles).
For traffic management applications, we decided to set the proximity vs. timeliness score to $1/9$ since the broadcasting decision does not have to be  space-dependent (i.e., \gls{ldm} updates should  be addressed to both spatially close and far neighbors). 
Attribute weights  $\vec{w}\hspace{-0.07cm}=\hspace{-0.07cm}\langle w_1,w_2,w_3 \rangle$ are determined from Eq.~\eqref{eq:eigenvector} and  demonstrate that the dissemination of space-related information is very valuable to safety services ($\max_{\vec{w}}\hspace{-0.07cm}=\hspace{-0.07cm}w_2\hspace{-0.07cm}=\hspace{-0.07cm}0.7471$) while, for traffic management services, time-related information should be preferred ($\max_{\vec{w}}\hspace{-0.07cm}=\hspace{-0.07cm}w_1\hspace{-0.07cm}=\hspace{-0.07cm}0.6554$).\hspace{0.05cm}
%Moreover, our conclusions have an empirical interpretation, since AHP assumes that the comparison process involves domain experts that assess which attributes and information sources are more important than others.

\begin{table}[t!]
\centering
%Different examples of reliable and consistent comparison score triplets ($\alpha,\beta,\gamma$) are considered, to model different interdependency characteristics among the considered attribute categories.}
%\vspace{-0.23cm}
\footnotesize
\centering
\begin{subtable}{.45\textwidth}
\centering
\begin{tabular}{c|ccc|c}
\Xhline{2\arrayrulewidth}
\multirow{2}{*}{Attribute}& \multicolumn{4}{c}{Application: Safety}             \\ \cline{2-5} 
& \begin{tabular}[c]{@{}c@{}}Timeliness\end{tabular} & \begin{tabular}[c]{@{}c@{}}Proximity\end{tabular}        & \begin{tabular}[c]{@{}c@{}}Quality\end{tabular}   & Weight  $\vec{w}$  \\\Xhline{2\arrayrulewidth}
\multicolumn{1}{l|}{Timeliness}        & 1               & 1/7        & 1 & \cellcolor{white!80!black}0.1194 \\ \cline{2-5} 
\multicolumn{1}{l|}{Proximity}      & 7      & 1                & 5  & \cellcolor{white!80!black}0.7471 \\ \cline{2-5} 
\multicolumn{1}{l|}{Quality}               & 1       & 1/5      & 1      & \cellcolor{white!80!black}0.1336 \\ \Xhline{2\arrayrulewidth}
\end{tabular}\vspace{0.23cm}
\end{subtable}\\ \vspace{0.1cm}
\begin{subtable}{.45\textwidth}
\centering
\begin{tabular}{c|ccc|c}
\Xhline{2\arrayrulewidth}
\multirow{2}{*}{Attribute}& \multicolumn{4}{c}{Application: Traffic Management}             \\ \cline{2-5} 
& \begin{tabular}[c]{@{}c@{}}Timeliness\end{tabular} & \begin{tabular}[c]{@{}c@{}}Proximity\end{tabular}        & \begin{tabular}[c]{@{}c@{}}Quality\end{tabular}   & Weight $\vec{w}$   \\\Xhline{2\arrayrulewidth}
\multicolumn{1}{l|}{Timeliness}        & 1               & 9       & 3 & \cellcolor{white!80!black}0.6554 \\ \cline{2-5} 
\multicolumn{1}{l|}{Proximity}      & 1/9      & 1                & 1/7  & \cellcolor{white!80!black}0.0549 \\ \cline{2-5} 
\multicolumn{1}{l|}{Quality}               & 1/3       & 7      & 1      & \cellcolor{white!80!black}0.2897 \\ \Xhline{2\arrayrulewidth}
\end{tabular}\vspace{0.0cm}
\end{subtable}
%\caption{ Parametric comparison matrix $M_a$ of \gls{voi} attributes for safety applications. The time-dependency attribute  is assumed to be at least ``moderately more important'' than the sensor's information quality (i.e.,  $\beta>3$, according to the Saaty scale \cite{saaty1990decision}).}
\vspace*{0.5cm}
\caption{Pairwise comparison matrices $M_a$ and  weights $\vec{w}=\langle w_1,w_2,w_3 \rangle$ of VoI attributes  for safety and traffic management applications.} 
\label{tab:attributes}
\vspace*{-0.8cm}
\end{table}

\begin{figure}[t!]
     \centering
         \begin{subfigure}[t!]{0.4\textwidth}
    \centering
  \setlength\fwidth{0.6\columnwidth}
  \setlength\fheight{0.5\columnwidth}
  \hspace{-0.66cm}% This file was created by matlab2tikz.
\pgfplotsset{
tick label style={font=\scriptsize},
label style={font=\scriptsize},
legend  style={font=\scriptsize}
}
\begin{tikzpicture}

\begin{axis}[%
every axis plot/.append style={ line width=0.7pt},
width=\fwidth,
height=\fheight,
at={(0\fwidth,0\fheight)},
scale only axis,
xmin=0,
xmax=500,
xlabel={TX-RX distance [m]},
ymin=0.2,
ymax=1,
ylabel={VoI},
ylabel style={font=\scriptsize\color{white!15!black}},
xlabel style={font=\scriptsize\color{white!15!black}},
axis background/.style={fill=white},
legend columns = {2},
xmajorgrids,
ymajorgrids,
yminorgrids,
legend style={at={(0.5,1.03)}, anchor=south, legend cell align=left, align=left, draw=white!15!black,/tikz/every even column/.append style={column sep=0.35cm}}
]
\addplot [color=red, mark=triangle,mark size=2.5pt, mark repeat = 6, mark options={solid, red},forget plot]
  table[row sep=crcr]{%
0 0.93742817831901\\
10  0.93709655437246\\
20  0.933733404744368\\
30  0.932356210294085\\
40  0.927412401900845\\
50  0.92403017120977\\
60  0.916983196184334\\
70  0.911833039160798\\
80  0.903708250152835\\
90  0.898311721225905\\
100 0.890689193440776\\
110 0.886322305850878\\
120 0.880003601897483\\
130 0.877002902796175\\
140 0.871970534740923\\
150 0.870100292849755\\
160 0.866015547434454\\
170 0.864913565295816\\
180 0.861436872611701\\
190 0.860807808248172\\
200 0.857694259536318\\
210 0.857341440597658\\
220 0.854436585549878\\
230 0.85424062542349\\
240 0.851453225789593\\
250 0.851344971438088\\
260 0.848622928358939\\
270 0.848563302236491\\
280 0.845877399977963\\
290 0.845844611536121\\
300 0.843178626256593\\
310 0.843160611985713\\
320 0.840505582289725\\
330 0.840495689941264\\
340 0.837846680312861\\
350 0.837841249496801\\
360 0.835195546017258\\
370 0.835192564985889\\
380 0.83254867664206\\
390 0.832547040455794\\
400 0.829904148485261\\
410 0.829903250478533\\
420 0.827260905391222\\
430 0.827260412540021\\
440 0.82461836760725\\
450 0.824618097120361\\
460 0.821976216921515\\
470 0.821976068473834\\
480 0.819334278684577\\
490 0.819334197214361\\
500 0.816692457043449\\
};

\addplot [color=red, dashed, mark=triangle,mark size=2.5pt, mark repeat = 6, mark options={solid, red}, forget plot]
  table[row sep=crcr]{%
0 0.937634030489672\\
10  0.937632244607895\\
20  0.93498433181558\\
30  0.934964074124465\\
40  0.932261985258109\\
50  0.932099152555745\\
60  0.929066106730102\\
70  0.928236295348592\\
80  0.924052143080171\\
90  0.921540618723192\\
100 0.915299882019928\\
110 0.910717349406356\\
120 0.902823924163254\\
130 0.897326391499071\\
140 0.889348968584133\\
150 0.884479318592679\\
160 0.877606060209122\\
170 0.874067574518417\\
180 0.868553609230066\\
190 0.866274513570453\\
200 0.861855369307624\\
210 0.860487063994692\\
220 0.856802305582444\\
230 0.856012951827084\\
240 0.852777180985193\\
250 0.852331867224073\\
260 0.849357402353736\\
270 0.849109269670863\\
280 0.846282885016835\\
290 0.846145564978445\\
300 0.843401887549276\\
310 0.843326178145853\\
320 0.840628330195229\\
330 0.840586675172411\\
340 0.837914112027509\\
350 0.837891219622243\\
360 0.835232573296709\\
370 0.835220000135312\\
380 0.832569003661972\\
390 0.832562100488409\\
400 0.829915306003478\\
410 0.829911516594851\\
420 0.827267029313253\\
430 0.827264949380621\\
440 0.824621728651448\\
450 0.824620587081571\\
460 0.821978061551238\\
470 0.821977435020717\\
480 0.81933529105376\\
490 0.819334947199384\\
500 0.816693012647932\\
};

\addplot [color=green!40!black, mark=o,mark size=1.5pt, mark repeat = 3, mark options={solid, green!40!black}, forget plot]
  table[row sep=crcr]{%
0 0.985828731646247\\
10  0.981316185092769\\
20  0.970281149630207\\
30  0.951541085365425\\
40  0.918997361109166\\
50  0.872973925205824\\
60  0.811811527792127\\
70  0.741731174524001\\
80  0.665902508818354\\
90  0.592469669413161\\
100 0.523475483465753\\
110 0.464053406899216\\
120 0.412800900343279\\
130 0.371969125290601\\
140 0.338220332456666\\
150 0.312771164240836\\
160 0.29191707132948\\
170 0.27692193676989\\
180 0.264341873176939\\
190 0.255781929737734\\
200 0.248143318948441\\
210 0.243342363220696\\
220 0.238543534914985\\
230 0.235877023035647\\
240 0.232676459959627\\
250 0.231203397456564\\
260 0.228892168516221\\
270 0.228080810783618\\
280 0.226261365193763\\
290 0.225815199071663\\
300 0.224266772194048\\
310 0.224021644431649\\
320 0.222622294788225\\
330 0.222487685441051\\
340 0.221170253389878\\
350 0.221096353991016\\
360 0.219823910098257\\
370 0.219783345950315\\
380 0.218535601372841\\
390 0.218513337097983\\
400 0.217279150591113\\
410 0.217266931035779\\
420 0.216040186198976\\
430 0.216033479765354\\
440 0.21481081925765\\
450 0.214807138628761\\
460 0.213586719724891\\
470 0.213584699734832\\
480 0.212365511072272\\
490 0.212364402472773\\
500 0.211145888987897\\
};

\addplot [color=green!40!black, dashed, mark=o,mark size=1.5pt, mark repeat = 3, mark options={solid, green!40!black}, forget plot]
  table[row sep=crcr]{%
0 0.988629848733554\\
10  0.988605547488902\\
20  0.987303040547808\\
30  0.987027385622486\\
40  0.984987682050276\\
50  0.982771948966191\\
60  0.9762287748168\\
70  0.964937182250511\\
80  0.942730418468723\\
90  0.908555049955834\\
100 0.858363476572918\\
110 0.796007027448307\\
120 0.723326625824875\\
130 0.648519386075025\\
140 0.574695992793464\\
150 0.508432616718976\\
160 0.449634054733161\\
170 0.401484391314058\\
180 0.361182302323464\\
190 0.330169688877388\\
200 0.304765290167559\\
210 0.286146184134351\\
220 0.270734882646336\\
230 0.25999381401943\\
240 0.250692075288403\\
250 0.244632503599269\\
260 0.238886465153479\\
270 0.235510019349269\\
280 0.231778970703501\\
290 0.229910399171181\\
300 0.227304782528641\\
310 0.226274572825651\\
320 0.224292577177933\\
330 0.223725759758364\\
340 0.222087825099319\\
350 0.221776318510132\\
360 0.22032775586697\\
370 0.220156667570364\\
380 0.218812199684272\\
390 0.218718265297459\\
400 0.217430975635412\\
410 0.217379411558036\\
420 0.21612351698237\\
430 0.21609521446402\\
440 0.21485655440003\\
450 0.214841020578459\\
460 0.213611820377557\\
470 0.213603294913177\\
480 0.212379286805645\\
490 0.212374607834496\\
500 0.211153449331846\\
};

\addplot [color=black]
  table[row sep=crcr]{%
  -1 0 \\
  };
  \addlegendentry{Urban}

  \addplot [color=black, dashed]
  table[row sep=crcr]{%
  -1 0 \\
  };
  \addlegendentry{Highway}

    \addplot [only marks, mark=o,mark size=1.5pt, mark repeat = 3, mark options={solid, green!40!black}]
  table[row sep=crcr]{%
  -1 0 \\
  };
  \addlegendentry{Safety}

      \addplot [only marks, mark=triangle,mark size=2.5pt, mark repeat = 3, mark options={solid, red}]
  table[row sep=crcr]{%
  -1 0 \\
  };
  \addlegendentry{Traffic Management}

  \node[coordinate] (B) at (axis cs:300,0.85) {};                       % for ellipse
\node[draw, coordinate, align=center, pin={[align=center,pin distance = 3mm, font=\scriptsize\linespread{0.8}\selectfont]-90:{No distinction between \\ urban and highway }}] at (axis cs:300,0.76){};  % for pin

\end{axis}
 \draw[black] (B) ellipse (0.2 and 0.3);                                % draw ellipse
\end{tikzpicture}%
  \vspace{-0.1cm}
    \caption{\footnotesize  Processed case. }
    \end{subfigure}\\\vspace{0.33cm}
      \begin{subfigure}[t!]{0.4\textwidth}
  \centering
  \setlength\fwidth{0.6\columnwidth}
  \setlength\fheight{0.5\columnwidth}
  \hspace{-0.66cm}% This file was created by matlab2tikz.
\pgfplotsset{
tick label style={font=\scriptsize},
label style={font=\scriptsize},
legend  style={font=\scriptsize}
}
\begin{tikzpicture}

\begin{axis}[%
every axis plot/.append style={ line width=0.7pt},
width=\fwidth,
height=\fheight,
at={(0\fwidth,0\fheight)},
scale only axis,
xmin=0,
xmax=500,
xlabel={TX-RX distance [m]},
ymin=0.1,
ymax=1,
ylabel={VoI},
ylabel style={font=\scriptsize\color{white!15!black}},
xlabel style={font=\scriptsize\color{white!15!black}},
axis background/.style={fill=white},
legend columns = {2},
xmajorgrids,
ymajorgrids,
yminorgrids,
legend style={at={(0.5,1.03)}, anchor=south, legend cell align=left, align=left, draw=white!15!black,/tikz/every even column/.append style={column sep=0.35cm}}
]
\addplot [color=red, mark=triangle,mark size=2.5pt, mark repeat = 6, mark options={solid, red},forget plot]
  table[row sep=crcr]{%
0 0.93742817831901\\
10  0.93709655437246\\
20  0.915608873427568\\
30  0.914231678977284\\
40  0.880741061109864\\
50  0.877358830418788\\
60  0.845363280025474\\
70  0.840213123001938\\
80  0.810324677376225\\
90  0.804928148449295\\
100 0.778360860769106\\
110 0.773993973179208\\
120 0.751225284303429\\
130 0.748224585202122\\
140 0.728950005792598\\
150 0.72707976390143\\
160 0.710706463315149\\
170 0.709604481176511\\
180 0.695567901768819\\
190 0.69493883740529\\
200 0.68279488523752\\
210 0.682442066298861\\
220 0.671859960134894\\
230 0.671664000008507\\
240 0.662396412405315\\
250 0.66228815805381\\
260 0.654144827833471\\
270 0.654085201711023\\
280 0.646914626676242\\
290 0.646881838234399\\
300 0.640559566778022\\
310 0.640541552507142\\
320 0.634962844283996\\
330 0.634952951935535\\
340 0.630028115119624\\
350 0.630022684303563\\
360 0.625674007937711\\
370 0.625671026906342\\
380 0.621830663123558\\
390 0.621829026937292\\
400 0.618437449075337\\
410 0.618436551068609\\
420 0.615441372757188\\
430 0.615440879905988\\
440 0.61279591101055\\
450 0.612795640523661\\
460 0.610460107095098\\
470 0.610459958647416\\
480 0.608397842920658\\
490 0.608397761450442\\
500 0.606577234310749\\
};

\addplot [color=red, dashed, mark=triangle,mark size=2.5pt, mark repeat = 6, mark options={solid, red}, forget plot]
  table[row sep=crcr]{%
0 0.937634030489672\\
10  0.937632244607895\\
20  0.934843688948168\\
30  0.934823431257054\\
40  0.926612742677946\\
50  0.926449909975582\\
60  0.918129038335292\\
70  0.917299226953782\\
80  0.908044692191742\\
90  0.905533167834762\\
100 0.894436161381827\\
110 0.889853628768254\\
120 0.87731471594235\\
130 0.871817183278166\\
140 0.859401724370217\\
150 0.854532074378762\\
160 0.843424901014905\\
170 0.8398864153242\\
180 0.830339325491182\\
190 0.82806022983157\\
200 0.819805420882628\\
210 0.818437115569697\\
220 0.811110821752812\\
230 0.810321467997452\\
240 0.803634960455322\\
250 0.803189646694203\\
260 0.796951913250946\\
270 0.796703780568072\\
280 0.790798264891364\\
290 0.790660944852974\\
300 0.785018943374286\\
310 0.784943233970863\\
320 0.779524538366803\\
330 0.779482883343985\\
340 0.774263618364651\\
350 0.774240725959385\\
360 0.769206193041343\\
370 0.769193619879946\\
380 0.764334221478946\\
390 0.764327318305382\\
400 0.75963627598056\\
410 0.759632486571933\\
420 0.755104574961132\\
430 0.755102495028499\\
440 0.750733342903734\\
450 0.750732201333857\\
460 0.746517906764464\\
470 0.746517280233943\\
480 0.742454199007378\\
490 0.742453855153003\\
500 0.738538484544318\\
};

\addplot [color=green!40!black, mark=o,mark size=1.5pt, mark repeat = 3, mark options={solid, green!40!black}, forget plot]
  table[row sep=crcr]{%
0 0.985828731646247\\
10  0.981316185092769\\
20  0.961926578984862\\
30  0.94318651472008\\
40  0.897484031636761\\
50  0.85146059573342\\
60  0.778798058990964\\
70  0.708717705722838\\
80  0.622857000924103\\
90  0.54942416151891\\
100 0.471697317488605\\
110 0.412275240922068\\
120 0.353440052157873\\
130 0.312608277105195\\
140 0.272294483564461\\
150 0.246845315348631\\
160 0.220326766870911\\
170 0.205331632311322\\
180 0.187883949352281\\
190 0.179324005913075\\
200 0.167522796802772\\
210 0.162721841075027\\
220 0.154384155091227\\
230 0.151717643211889\\
240 0.145530013202204\\
250 0.144056950699141\\
260 0.139246759130848\\
270 0.138435401398246\\
280 0.134548729082603\\
290 0.134102562960503\\
300 0.130868757265077\\
310 0.130623629502678\\
320 0.127876599283548\\
330 0.127741989936374\\
340 0.125375506812802\\
350 0.12530160741394\\
360 0.123244171786107\\
370 0.123203607638164\\
380 0.121404343222442\\
390 0.121382078947584\\
400 0.119802782872818\\
410 0.119790563317483\\
420 0.118401178683529\\
430 0.118394472249906\\
440 0.117170463930651\\
450 0.117166783301762\\
460 0.116087576089912\\
470 0.116085556099853\\
480 0.115133570369152\\
490 0.115132461769652\\
500 0.114292489509611\\
};

\addplot [color=green!40!black, dashed, mark=o,mark size=1.5pt, mark repeat = 3, mark options={solid, green!40!black}, forget plot]
  table[row sep=crcr]{%
0 0.988629848733554\\
10  0.988605547488902\\
20  0.987238210690913\\
30  0.986962555765592\\
40  0.982383642508303\\
50  0.980167909424219\\
60  0.971187292145592\\
70  0.959895699579303\\
80  0.935351723982134\\
90  0.901176355469245\\
100 0.848746266342814\\
110 0.786389817218203\\
120 0.711568060681134\\
130 0.636760820931284\\
140 0.560891698323976\\
150 0.494628322249487\\
160 0.433878121283827\\
170 0.385728457864725\\
180 0.343567284998197\\
190 0.312554671552121\\
200 0.285382208828286\\
210 0.266763102795078\\
220 0.249673221912992\\
230 0.238932153286087\\
240 0.228039784538938\\
250 0.221980212849804\\
260 0.214729958523854\\
270 0.211353512719644\\
280 0.206203127087688\\
290 0.204334555555368\\
300 0.200392945578624\\
310 0.199362735875635\\
320 0.196126555303709\\
330 0.19555973788414\\
340 0.192747891468896\\
350 0.192436384879709\\
360 0.189892648406367\\
370 0.189721560109761\\
380 0.187359121077521\\
390 0.187265186690709\\
400 0.185035593324557\\
410 0.184984029247182\\
420 0.182859963167466\\
430 0.182831660649116\\
440 0.180797426039145\\
450 0.180781892217574\\
460 0.178828179186769\\
470 0.178819653722389\\
480 0.176940659259044\\
490 0.176935980287896\\
500 0.175127826661537\\
};

% \addplot [color=black]
%   table[row sep=crcr]{%
%   -1 0 \\
%   };
%   \addlegendentry{Urban}

%   \addplot [color=black, dashed]
%   table[row sep=crcr]{%
%   -1 0 \\
%   };
%   \addlegendentry{Highway}

%     \addplot [only marks, mark=o,mark size=1.5pt, mark repeat = 3, mark options={solid, green!40!black}]
%   table[row sep=crcr]{%
%   -1 0 \\
%   };
%   \addlegendentry{Safety}

%       \addplot [only marks, mark=triangle,mark size=2.5pt, mark repeat = 3, mark options={solid, red}]
%   table[row sep=crcr]{%
%   -1 0 \\
%   };
%   \addlegendentry{Traffic Management}

\node[coordinate] (B) at (axis cs:363,0.7) {};                       % for ellipse
\node[draw, coordinate, align=center, pin={[align=center,pin distance = 3mm, font=\scriptsize\linespread{0.8}\selectfont]-90:{Large gap between \\ urban and highway }}] at (axis cs:363,0.55){};  % for pin

\end{axis}
\draw[black] (B) ellipse (0.2 and 0.6);                                % draw ellipse
\end{tikzpicture}%
  \vspace{-0.1cm}
    \caption{\footnotesize  Non-processed case.}
    \end{subfigure}
    \setlength{\belowcaptionskip}{-0.6cm}
    \caption{\footnotesize  VoI vs. TX-RX distance for urban and highway scenarios for processed and non-processed strategies. $P_{\rm td}=1$, $r_h=1080$ px, AoI 
  ~$=100$~ms.}
    \label{fig:voi_scenario}
    \vspace*{0.15cm}
\end{figure}
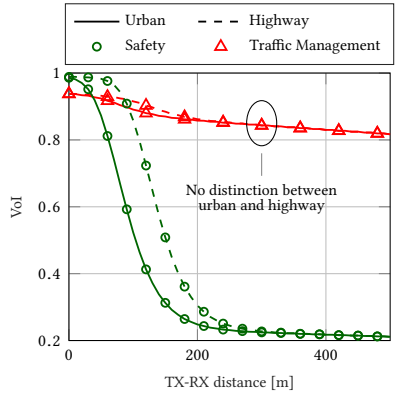
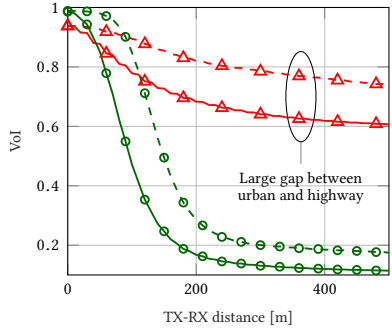

\smallskip
\textbf{Phase 2 results -- \emph{conditional VoI}.}
In Fig.~\ref{fig:voi_attributes}a we plot the conditional VoI for the proximity attribute as defined by the logistic function in Eq.~\eqref{eq:proximity}.
We can see that the value of position information is generally higher for a highway scenario than for an urban one. This is consistent with the fact that the larger safety distance between vehicles in highway scenarios requires the information to be disseminated over longer ranges, in order to reach the nearby vehicles. Furthermore, the highway scenario is usually characterized by better propagation conditions that increase the probability of successful reception at long distances and, consequently, the value of the packet transmission itself. Accordingly,  VoI drops to zero beyond 300 m, which thus represents a suitable communication range for vehicular networks.
In Fig.~\ref{fig:voi_attributes}b we plot the conditional VoI for the timeliness attribute as follows from the exponential function in Eq.~\eqref{eq:timeliness}, which is proportional to the \gls{aoi} and the  temporal characteristics of the perceived object, as defined in Sec.~\ref{sub:system_model}.
In particular,  the value is constant in case of static observations while,\hspace{0.05cm}for\hspace{0.05cm}dynamic\hspace{0.05cm}observations, it\hspace{0.05cm}decreases at a pace\hspace{0.05cm}that\hspace{0.05cm}is\hspace{0.05cm}a\hspace{0.05cm}function of~$P_{\rm td}$.

In Fig.~\ref{fig:voi_attributes}c and  Fig.~\ref{fig:voi_attributes}d we plot the conditional VoI for the quality attribute, considering both processed and non-processed value assessment strategies, respectively. 
In the non-processed case the conditional VoI exhibits a significant difference between urban and highway scenarios. This gap is due to the higher probability that the line of sight is blocked in urban scenarios, in which case the image captured by the camera would be basically useless.  
%In both cases, inaccurate camera electronics make the conditional VoI  drop at large distance, thereby calling for short-range broadcasting.

%In the first case,  the value depends on the resolution of the sensor which, in turn, determines the level of detection detail, i.e., accuracy.
%While, for high-quality sensors (i.e., $4096\times780$~px cameras, corresponding to a focal distance of 773 mm according to Eq.~\eqref{eq:fd}), the conditional VoI is not significantly affected by the observation distance, for low-quality sensors (i.e., $640\times480$ px cameras, corresponding to a focal distance of 120 mm according to Eq.~\eqref{eq:fd}), inaccurate camera electronics make the conditional VoI  drop at large distance, thereby calling for short-range broadcasting operations.
%In the second case, the sender does not access the perception record  to determine, prior to transmission, whether obstacles in the path of motion prevented object detection. 
%Detection capabilities are thus modeled as a function of the \gls{los} probability, as described in Eq.~\eqref{eq:PL_LOS} which, at large distance, put an additional strain  on already degraded observations. 

\smallskip
\textbf{Phase 3 results -- \emph{overall VoI}.}
Our goal is to assign a value to different sources of information (in this work we consider camera observations) in such a way that the utility to potential receiver(s) is maximized.
In Fig.~\ref{fig:voi_scenario} we evaluate the impact of the propagation scenario on  VoI.
First, we observe that, for safety applications, the value of data transmission at short distances is high in all considered conditions, reflecting the importance of maintaining fresh and updated information among close-by vehicles.
% we observe that, at short distance,  \gls{voi} is higher considering safety applications in all investigated configurations because the potentially detrimental effects deriving from discontinuous communications would call for persistent dissemination of sensory information.
Second, we see that, for traffic management services,  \gls{voi} is almost independent of the TX/RX distance (proximity weight $w_2=0.0549\ll1$), demonstrating the importance of sharing \gls{ldm} updates even at large distances. 
%Minor VoI decrease is due to information quality degradation as a function of the distance, as illustrated in Fig.~\ref{fig:voi_attributes}.
Third, we notice that, for safety services,  VoI drops for values of $d$ larger than 100 m, which is a rather safe inter-vehicle distance (in particular, in urban scenarios):  transmitting data beyond this range would just increase the channel access contention without bringing much benefit in terms of safety.
We can also observe that, in general,  VoI for the considered information is higher in highway than in urban scenarios, because of the higher probability of \gls{los} between object and sensor (camera).
In general, the non-processed case (Fig.~\ref{fig:voi_scenario}b), although guaranteeing real-time value determination, represents a lower bound for  VoI in vehicular networks, as it provides a probabilistic, rather than deterministic, method to  assess  VoI.
Moreover, for the non-processed case, different characterizations of the quality attribute in urban and highway scenarios (see Fig.~\ref{fig:voi_attributes}d) result in different overall VoI: for traffic management applications, the gap is as large as 25\% when $d>200$ m, i.e., when the endpoints are in~NLOS.

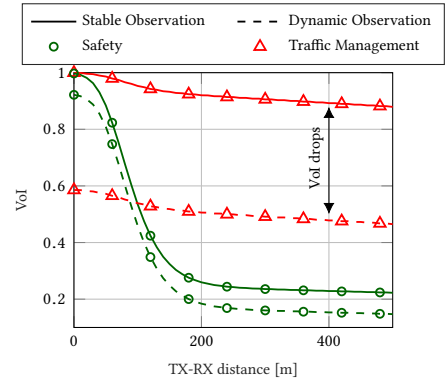
\begin{figure}[t!]
     \centering
  \setlength\fwidth{0.5\columnwidth}
  \setlength\fheight{0.4\columnwidth}
\pgfplotsset{
tick label style={font=\scriptsize},
label style={font=\scriptsize},
legend  style={font=\scriptsize}
}
\begin{tikzpicture}

\begin{axis}[%
every axis plot/.append style={ line width=0.7pt},
width=\fwidth,
height=\fheight,
at={(0\fwidth,0\fheight)},
scale only axis,
xmin=0,
xmax=500,
xlabel={TX-RX distance [m]},
ymin=0.1,
ymax=1,
ylabel={VoI},
ylabel style={font=\scriptsize\color{white!15!black}},
xlabel style={font=\scriptsize\color{white!15!black}},
axis background/.style={fill=white},
legend columns = {2},
xmajorgrids,
ymajorgrids,
yminorgrids,
legend style={at={(0.5,1.03)}, anchor=south, legend cell align=left, align=left, draw=white!15!black,/tikz/every even column/.append style={column sep=0.35cm}}
]
\addplot [color=red, mark=triangle,mark size=2.5pt, mark repeat = 6, mark options={solid, red},forget plot]
  table[row sep=crcr]{%
0 0.999793498914539\\
10  0.999461874967989\\
20  0.996098725339897\\
30  0.994721530889613\\
40  0.989777722496374\\
50  0.986395491805299\\
60  0.979348516779863\\
70  0.974198359756327\\
80  0.966073570748363\\
90  0.960677041821434\\
100 0.953054514036305\\
110 0.948687626446407\\
120 0.942368922493012\\
130 0.939368223391704\\
140 0.934335855336452\\
150 0.932465613445284\\
160 0.928380868029983\\
170 0.927278885891345\\
180 0.923802193207229\\
190 0.923173128843701\\
200 0.920059580131846\\
210 0.919706761193187\\
220 0.916801906145406\\
230 0.916605946019019\\
240 0.913818546385121\\
250 0.913710292033617\\
260 0.910988248954467\\
270 0.910928622832019\\
280 0.908242720573492\\
290 0.90820993213165\\
300 0.905543946852122\\
310 0.905525932581242\\
320 0.902870902885254\\
330 0.902861010536793\\
340 0.90021200090839\\
350 0.900206570092329\\
360 0.897560866612786\\
370 0.897557885581417\\
380 0.894913997237589\\
390 0.894912361051323\\
400 0.89226946908079\\
410 0.892268571074062\\
420 0.88962622598675\\
430 0.88962573313555\\
440 0.886983688202779\\
450 0.88698341771589\\
460 0.884341537517044\\
470 0.884341389069362\\
480 0.881699599280106\\
490 0.88169951780989\\
500 0.879057777638977\\
};

\addplot [color=green!40!black, mark=o,mark size=1.5pt, mark repeat = 6, mark options={solid, green!40!black},forget plot]
  table[row sep=crcr]{%
0 0.997190052856017\\
10  0.992677506302539\\
20  0.981642470839977\\
30  0.962902406575195\\
40  0.930358682318935\\
50  0.884335246415594\\
60  0.823172849001897\\
70  0.753092495733771\\
80  0.677263830028124\\
90  0.603830990622931\\
100 0.534836804675523\\
110 0.475414728108986\\
120 0.424162221553049\\
130 0.383330446500371\\
140 0.349581653666436\\
150 0.324132485450605\\
160 0.30327839253925\\
170 0.28828325797966\\
180 0.275703194386709\\
190 0.267143250947504\\
200 0.259504640158211\\
210 0.254703684430466\\
220 0.249904856124755\\
230 0.247238344245417\\
240 0.244037781169397\\
250 0.242564718666334\\
260 0.24025348972599\\
270 0.239442131993388\\
280 0.237622686403533\\
290 0.237176520281433\\
300 0.235628093403818\\
310 0.235382965641419\\
320 0.233983615997994\\
330 0.233849006650821\\
340 0.232531574599648\\
350 0.232457675200786\\
360 0.231185231308027\\
370 0.231144667160085\\
380 0.229896922582611\\
390 0.229874658307753\\
400 0.228640471800883\\
410 0.228628252245548\\
420 0.227401507408746\\
430 0.227394800975124\\
440 0.22617214046742\\
450 0.22616845983853\\
460 0.224948040934661\\
470 0.224946020944602\\
480 0.223726832282042\\
490 0.223725723682542\\
500 0.222507210197667\\
};

\addplot [color=red, dashed, mark=triangle,mark size=2.5pt, mark repeat = 6, mark options={solid, red},forget plot]
  table[row sep=crcr]{%
0 0.585529819927093\\
10  0.585198195980543\\
20  0.581835046352451\\
30  0.580457851902167\\
40  0.575514043508928\\
50  0.572131812817853\\
60  0.565084837792417\\
70  0.559934680768881\\
80  0.551809891760917\\
90  0.546413362833988\\
100 0.538790835048859\\
110 0.534423947458961\\
120 0.528105243505566\\
130 0.525104544404258\\
140 0.520072176349006\\
150 0.518201934457838\\
160 0.514117189042537\\
170 0.513015206903899\\
180 0.509538514219784\\
190 0.508909449856255\\
200 0.5057959011444\\
210 0.505443082205741\\
220 0.502538227157961\\
230 0.502342267031573\\
240 0.499554867397676\\
250 0.499446613046171\\
260 0.496724569967022\\
270 0.496664943844574\\
280 0.493979041586046\\
290 0.493946253144204\\
300 0.491280267864676\\
310 0.491262253593796\\
320 0.488607223897808\\
330 0.488597331549347\\
340 0.485948321920944\\
350 0.485942891104883\\
360 0.48329718762534\\
370 0.483294206593972\\
380 0.480650318250143\\
390 0.480648682063877\\
400 0.478005790093344\\
410 0.478004892086616\\
420 0.475362546999304\\
430 0.475362054148104\\
440 0.472720009215333\\
450 0.472719738728444\\
460 0.470077858529598\\
470 0.470077710081916\\
480 0.46743592029266\\
490 0.467435838822444\\
500 0.464794098651532\\
};

\addplot [color=green!40!black, dashed, mark=o,mark size=1.5pt, mark repeat = 6, mark options={solid, green!40!black},forget plot]
  table[row sep=crcr]{%
0 0.921722105644746\\
10  0.917209559091268\\
20  0.906174523628706\\
30  0.887434459363923\\
40  0.854890735107664\\
50  0.808867299204323\\
60  0.747704901790626\\
70  0.6776245485225\\
80  0.601795882816852\\
90  0.52836304341166\\
100 0.459368857464252\\
110 0.399946780897714\\
120 0.348694274341778\\
130 0.3078624992891\\
140 0.274113706455165\\
150 0.248664538239334\\
160 0.227810445327979\\
170 0.212815310768389\\
180 0.200235247175438\\
190 0.191675303736233\\
200 0.184036692946939\\
210 0.179235737219195\\
220 0.174436908913484\\
230 0.171770397034146\\
240 0.168569833958126\\
250 0.167096771455063\\
260 0.164785542514719\\
270 0.163974184782117\\
280 0.162154739192262\\
290 0.161708573070161\\
300 0.160160146192547\\
310 0.159915018430148\\
320 0.158515668786723\\
330 0.158381059439549\\
340 0.157063627388377\\
350 0.156989727989515\\
360 0.155717284096756\\
370 0.155676719948814\\
380 0.15442897537134\\
390 0.154406711096482\\
400 0.153172524589612\\
410 0.153160305034277\\
420 0.151933560197475\\
430 0.151926853763853\\
440 0.150704193256149\\
450 0.150700512627259\\
460 0.14948009372339\\
470 0.149478073733331\\
480 0.148258885070771\\
490 0.148257776471271\\
500 0.147039262986396\\
};

\addplot [color=black]
  table[row sep=crcr]{%
  -1 0 \\
  };
  \addlegendentry{Stable Observation}

  \addplot [color=black, dashed]
  table[row sep=crcr]{%
  -1 0 \\
  };
  \addlegendentry{Dynamic Observation}

    \addplot [only marks, mark=o,mark size=1.5pt, mark repeat = 3, mark options={solid, green!40!black}]
  table[row sep=crcr]{%
  -1 0 \\
  };
  \addlegendentry{Safety}

      \addplot [only marks, mark=triangle,mark size=2.5pt, mark repeat = 3, mark options={solid, red}]
  table[row sep=crcr]{%
  -1 0 \\
  };
  \addlegendentry{Traffic Management}

\draw[{Latex[length=1.5mm,width=1.5mm]}-{Latex[length=1.5mm,width=1.5mm]}] (400,0.5)--(400,0.88);
\node[align=right, rotate=90] at (380,0.7) {\scriptsize VoI drops};
%   \node[coordinate] (B) at (axis cs:40,0.93) {};                       % for ellipse
% \node[draw, coordinate, align=center, pin={[align=left,pin distance = 5mm, font=\scriptsize\linespread{0.8}\selectfont]-0:{ Safety app \\  more valuable}}] at (axis cs:120,0.93){};  % for pin

\end{axis}
%\draw[black] (B) ellipse (0.6 and 0.2);                                % draw ellipse
\end{tikzpicture}%
    \vspace*{-0.33cm}
    \caption{\footnotesize VoI vs. TX-RX distance for\hspace{0.05cm}different types of observations. Processed VoI operations, $r_h =\hspace{-0.07cm}1080$ px, AoI $\hspace{-0.07cm}=\hspace{-0.07cm}100$ ms, urban scenarios are considered.  }
    \label{fig:voi_observation}
    \vspace{0.15cm}
\end{figure}

  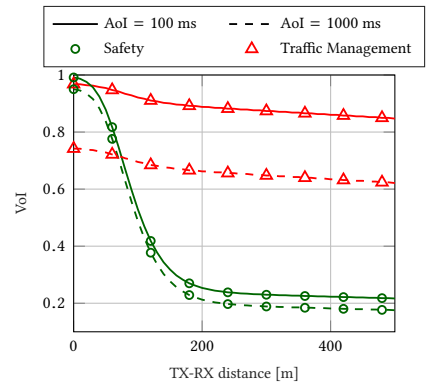
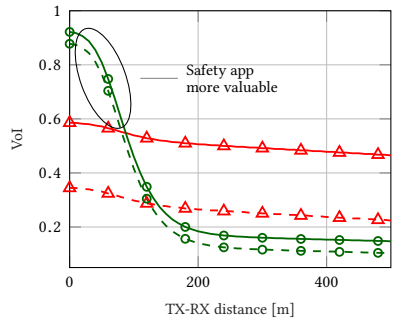
\begin{figure}[t!]
     \centering
                     \begin{subfigure}[t!]{0.4\textwidth}
                     \centering
          \setlength{\belowcaptionskip}{0cm}
  \setlength{\belowcaptionskip}{0cm}
    \setlength\fwidth{0.6\columnwidth}
  \setlength\fheight{0.48\columnwidth}
\pgfplotsset{
tick label style={font=\scriptsize},
label style={font=\scriptsize},
legend  style={font=\scriptsize}
}
\begin{tikzpicture}

\begin{axis}[%
every axis plot/.append style={ line width=0.7pt},
width=\fwidth,
height=\fheight,
at={(0\fwidth,0\fheight)},
scale only axis,
xmin=0,
xmax=500,
xlabel={TX-RX distance [m]},
ymin=0.1,
ymax=1,
ylabel={VoI},
ylabel style={font=\scriptsize\color{white!15!black}},
xlabel style={font=\scriptsize\color{white!15!black}},
axis background/.style={fill=white},
legend columns = {2},
xmajorgrids,
ymajorgrids,
yminorgrids,
legend style={at={(0.5,1.03)}, anchor=south, legend cell align=left, align=left, draw=white!15!black,/tikz/every even column/.append style={column sep=0.35cm}}
]
\addplot [color=red, mark=triangle,mark size=2.5pt, mark repeat = 6, mark options={solid, red},forget plot]
  table[row sep=crcr]{%
0 0.967831434478428\\
10  0.967499810531877\\
20  0.964136660903785\\
30  0.962759466453502\\
40  0.957815658060262\\
50  0.954433427369187\\
60  0.947386452343751\\
70  0.942236295320215\\
80  0.934111506312252\\
90  0.928714977385322\\
100 0.921092449600194\\
110 0.916725562010296\\
120 0.9104068580569\\
130 0.907406158955593\\
140 0.90237379090034\\
150 0.900503549009172\\
160 0.896418803593872\\
170 0.895316821455233\\
180 0.891840128771118\\
190 0.891211064407589\\
200 0.888097515695735\\
210 0.887744696757075\\
220 0.884839841709295\\
230 0.884643881582907\\
240 0.88185648194901\\
250 0.881748227597505\\
260 0.879026184518356\\
270 0.878966558395908\\
280 0.87628065613738\\
290 0.876247867695538\\
300 0.873581882416011\\
310 0.87356386814513\\
320 0.870908838449142\\
330 0.870898946100681\\
340 0.868249936472279\\
350 0.868244505656218\\
360 0.865598802176675\\
370 0.865595821145306\\
380 0.862951932801477\\
390 0.862950296615211\\
400 0.860307404644678\\
410 0.86030650663795\\
420 0.857664161550639\\
430 0.857663668699438\\
440 0.855021623766667\\
450 0.855021353279778\\
460 0.852379473080933\\
470 0.852379324633251\\
480 0.849737534843994\\
490 0.849737453373778\\
500 0.847095713202866\\
};

\addplot [color=green!40!black, mark=o,mark size=1.5pt, mark repeat = 6, mark options={solid, green!40!black},forget plot]
  table[row sep=crcr]{%
0 0.991367405315389\\
10  0.986854858761911\\
20  0.975819823299349\\
30  0.957079759034567\\
40  0.924536034778308\\
50  0.878512598874966\\
60  0.817350201461269\\
70  0.747269848193143\\
80  0.671441182487496\\
90  0.598008343082303\\
100 0.529014157134895\\
110 0.469592080568358\\
120 0.418339574012421\\
130 0.377507798959743\\
140 0.343759006125808\\
150 0.318309837909978\\
160 0.297455744998622\\
170 0.282460610439032\\
180 0.269880546846081\\
190 0.261320603406876\\
200 0.253681992617583\\
210 0.248881036889838\\
220 0.244082208584127\\
230 0.241415696704789\\
240 0.238215133628769\\
250 0.236742071125706\\
260 0.234430842185363\\
270 0.23361948445276\\
280 0.231800038862905\\
290 0.231353872740805\\
300 0.22980544586319\\
310 0.229560318100791\\
320 0.228160968457367\\
330 0.228026359110193\\
340 0.22670892705902\\
350 0.226635027660158\\
360 0.225362583767399\\
370 0.225322019619457\\
380 0.224074275041983\\
390 0.224052010767125\\
400 0.222817824260255\\
410 0.222805604704921\\
420 0.221578859868118\\
430 0.221572153434496\\
440 0.220349492926792\\
450 0.220345812297903\\
460 0.219125393394033\\
470 0.219123373403974\\
480 0.217904184741414\\
490 0.217903076141915\\
500 0.216684562657039\\
};

\addplot [color=red, dashed, mark=triangle,mark size=2.5pt, mark repeat = 6, mark options={solid, red},forget plot]
  table[row sep=crcr]{%
0 0.741931206350794\\
10  0.741599582404243\\
20  0.738236432776152\\
30  0.736859238325868\\
40  0.731915429932629\\
50  0.728533199241553\\
60  0.721486224216118\\
70  0.716336067192581\\
80  0.708211278184618\\
90  0.702814749257689\\
100 0.69519222147256\\
110 0.690825333882662\\
120 0.684506629929267\\
130 0.681505930827959\\
140 0.676473562772707\\
150 0.674603320881539\\
160 0.670518575466238\\
170 0.6694165933276\\
180 0.665939900643484\\
190 0.665310836279955\\
200 0.662197287568101\\
210 0.661844468629442\\
220 0.658939613581661\\
230 0.658743653455274\\
240 0.655956253821376\\
250 0.655847999469872\\
260 0.653125956390722\\
270 0.653066330268274\\
280 0.650380428009746\\
290 0.650347639567904\\
300 0.647681654288377\\
310 0.647663640017496\\
320 0.645008610321508\\
330 0.644998717973047\\
340 0.642349708344645\\
350 0.642344277528584\\
360 0.639698574049041\\
370 0.639695593017672\\
380 0.637051704673843\\
390 0.637050068487577\\
400 0.634407176517044\\
410 0.634406278510316\\
420 0.631763933423005\\
430 0.631763440571805\\
440 0.629121395639033\\
450 0.629121125152144\\
460 0.626479244953299\\
470 0.626479096505617\\
480 0.62383730671636\\
490 0.623837225246144\\
500 0.621195485075232\\
};

\addplot [color=green!40!black, dashed, mark=o,mark size=1.5pt, mark repeat = 6, mark options={solid, green!40!black},forget plot]
  table[row sep=crcr]{%
0 0.950214324907712\\
10  0.945701778354234\\
20  0.934666742891672\\
30  0.91592667862689\\
40  0.883382954370631\\
50  0.83735951846729\\
60  0.776197121053592\\
70  0.706116767785466\\
80  0.630288102079819\\
90  0.556855262674626\\
100 0.487861076727218\\
110 0.428439000160681\\
120 0.377186493604744\\
130 0.336354718552066\\
140 0.302605925718131\\
150 0.277156757502301\\
160 0.256302664590945\\
170 0.241307530031356\\
180 0.228727466438404\\
190 0.220167522999199\\
200 0.212528912209906\\
210 0.207727956482162\\
220 0.202929128176451\\
230 0.200262616297112\\
240 0.197062053221092\\
250 0.195588990718029\\
260 0.193277761777686\\
270 0.192466404045083\\
280 0.190646958455229\\
290 0.190200792333128\\
300 0.188652365455513\\
310 0.188407237693114\\
320 0.18700788804969\\
330 0.186873278702516\\
340 0.185555846651344\\
350 0.185481947252482\\
360 0.184209503359722\\
370 0.18416893921178\\
380 0.182921194634306\\
390 0.182898930359448\\
400 0.181664743852578\\
410 0.181652524297244\\
420 0.180425779460441\\
430 0.180419073026819\\
440 0.179196412519115\\
450 0.179192731890226\\
460 0.177972312986356\\
470 0.177970292996297\\
480 0.176751104333737\\
490 0.176749995734238\\
500 0.175531482249362\\
};

\addplot [color=black]
  table[row sep=crcr]{%
  -1 0 \\
  };
  \addlegendentry{$\text{AoI}=100$ ms}

  \addplot [color=black, dashed]
  table[row sep=crcr]{%
  -1 0 \\
  };
  \addlegendentry{$\text{AoI}=1000$ ms}

    \addplot [only marks, mark=o,mark size=1.5pt, mark repeat = 3, mark options={solid, green!40!black}]
  table[row sep=crcr]{%
  -1 0 \\
  };
  \addlegendentry{Safety}

      \addplot [only marks, mark=triangle,mark size=2.5pt, mark repeat = 3, mark options={solid, red}]
  table[row sep=crcr]{%
  -1 0 \\
  };
  \addlegendentry{Traffic Management}

% \draw[{Latex[length=1.5mm,width=1.5mm]}-{Latex[length=1.5mm,width=1.5mm]}] (400,0.26)--(400,0.67);
% \node[align=right, rotate=90] at (380,0.47) {\scriptsize VoI drops};
%   \node[coordinate] (B) at (axis cs:40,0.93) {};                       % for ellipse
% \node[draw, coordinate, align=center, pin={[align=left,pin distance = 5mm, font=\scriptsize\linespread{0.8}\selectfont]-0:{ Safety app \\  more valuable}}] at (axis cs:120,0.93){};  % for pin

\end{axis}
%\draw[black] (B) ellipse (0.6 and 0.2);                                % draw ellipse
\end{tikzpicture}%
  \vspace{-0.1cm}
    \caption{\footnotesize  Static observations.}
    \label{fig:voi_aoi_stable}
      \end{subfigure} \\\vspace{0.05cm}
                   \begin{subfigure}[t!]{0.4\textwidth}
                   \centering
    \setlength\fwidth{0.6\columnwidth}
  \setlength\fheight{0.48\columnwidth}
 \hspace{-0.66cm} % This file was created by matlab2tikz.
\pgfplotsset{
tick label style={font=\scriptsize},
label style={font=\scriptsize},
legend  style={font=\scriptsize}
}
\begin{tikzpicture}

\begin{axis}[%
every axis plot/.append style={ line width=0.7pt},
width=\fwidth,
height=\fheight,
at={(0\fwidth,0\fheight)},
scale only axis,
xmin=0,
xmax=500,
xlabel={TX-RX distance [m]},
ymin=0.05,
ymax=1,
ylabel={VoI},
ylabel style={font=\scriptsize\color{white!15!black}},
xlabel style={font=\scriptsize\color{white!15!black}},
axis background/.style={fill=white},
legend columns = {2},
xmajorgrids,
ymajorgrids,
yminorgrids,
legend style={at={(0.5,1.03)}, anchor=south, legend cell align=left, align=left, draw=white!15!black,/tikz/every even column/.append style={column sep=0.35cm}}
]
\addplot [color=red, mark=triangle,mark size=2.5pt, mark repeat = 6, mark options={solid, red},forget plot]
  table[row sep=crcr]{%
0 0.585529819927093\\
10  0.585198195980543\\
20  0.581835046352451\\
30  0.580457851902167\\
40  0.575514043508928\\
50  0.572131812817853\\
60  0.565084837792417\\
70  0.559934680768881\\
80  0.551809891760917\\
90  0.546413362833988\\
100 0.538790835048859\\
110 0.534423947458961\\
120 0.528105243505566\\
130 0.525104544404258\\
140 0.520072176349006\\
150 0.518201934457838\\
160 0.514117189042537\\
170 0.513015206903899\\
180 0.509538514219784\\
190 0.508909449856255\\
200 0.5057959011444\\
210 0.505443082205741\\
220 0.502538227157961\\
230 0.502342267031573\\
240 0.499554867397676\\
250 0.499446613046171\\
260 0.496724569967022\\
270 0.496664943844574\\
280 0.493979041586046\\
290 0.493946253144204\\
300 0.491280267864676\\
310 0.491262253593796\\
320 0.488607223897808\\
330 0.488597331549347\\
340 0.485948321920944\\
350 0.485942891104883\\
360 0.48329718762534\\
370 0.483294206593972\\
380 0.480650318250143\\
390 0.480648682063877\\
400 0.478005790093344\\
410 0.478004892086616\\
420 0.475362546999304\\
430 0.475362054148104\\
440 0.472720009215333\\
450 0.472719738728444\\
460 0.470077858529598\\
470 0.470077710081916\\
480 0.46743592029266\\
490 0.467435838822444\\
500 0.464794098651532\\
};

\addplot [color=green!40!black, mark=o,mark size=1.5pt, mark repeat = 6, mark options={solid, green!40!black},forget plot]
  table[row sep=crcr]{%
0 0.921722105644746\\
10  0.917209559091268\\
20  0.906174523628706\\
30  0.887434459363923\\
40  0.854890735107664\\
50  0.808867299204323\\
60  0.747704901790626\\
70  0.6776245485225\\
80  0.601795882816852\\
90  0.52836304341166\\
100 0.459368857464252\\
110 0.399946780897714\\
120 0.348694274341778\\
130 0.3078624992891\\
140 0.274113706455165\\
150 0.248664538239334\\
160 0.227810445327979\\
170 0.212815310768389\\
180 0.200235247175438\\
190 0.191675303736233\\
200 0.184036692946939\\
210 0.179235737219195\\
220 0.174436908913484\\
230 0.171770397034146\\
240 0.168569833958126\\
250 0.167096771455063\\
260 0.164785542514719\\
270 0.163974184782117\\
280 0.162154739192262\\
290 0.161708573070161\\
300 0.160160146192547\\
310 0.159915018430148\\
320 0.158515668786723\\
330 0.158381059439549\\
340 0.157063627388377\\
350 0.156989727989515\\
360 0.155717284096756\\
370 0.155676719948814\\
380 0.15442897537134\\
390 0.154406711096482\\
400 0.153172524589612\\
410 0.153160305034277\\
420 0.151933560197475\\
430 0.151926853763853\\
440 0.150704193256149\\
450 0.150700512627259\\
460 0.14948009372339\\
470 0.149478073733331\\
480 0.148258885070771\\
490 0.148257776471271\\
500 0.147039262986396\\
};

\addplot [color=red, dashed, mark=triangle,mark size=2.5pt, mark repeat = 6, mark options={solid, red},forget plot]
  table[row sep=crcr]{%
0 0.344467761347653\\
10  0.344136137401103\\
20  0.340772987773011\\
30  0.339395793322727\\
40  0.334451984929488\\
50  0.331069754238413\\
60  0.324022779212977\\
70  0.318872622189441\\
80  0.310747833181477\\
90  0.305351304254548\\
100 0.297728776469419\\
110 0.293361888879521\\
120 0.287043184926126\\
130 0.284042485824818\\
140 0.279010117769566\\
150 0.277139875878398\\
160 0.273055130463097\\
170 0.271953148324459\\
180 0.268476455640344\\
190 0.267847391276815\\
200 0.26473384256496\\
210 0.264381023626301\\
220 0.261476168578521\\
230 0.261280208452133\\
240 0.258492808818236\\
250 0.258384554466731\\
260 0.255662511387582\\
270 0.255602885265134\\
280 0.252916983006606\\
290 0.252884194564764\\
300 0.250218209285236\\
310 0.250200195014356\\
320 0.247545165318368\\
330 0.247535272969907\\
340 0.244886263341504\\
350 0.244880832525443\\
360 0.2422351290459\\
370 0.242232148014532\\
380 0.239588259670703\\
390 0.239586623484437\\
400 0.236943731513904\\
410 0.236942833507176\\
420 0.234300488419864\\
430 0.234299995568664\\
440 0.231657950635893\\
450 0.231657680149004\\
460 0.229015799950158\\
470 0.229015651502476\\
480 0.22637386171322\\
490 0.226373780243004\\
500 0.223732040072091\\
};

\addplot [color=green!40!black, dashed, mark=o,mark size=1.5pt, mark repeat = 6, mark options={solid, green!40!black},forget plot]
  table[row sep=crcr]{%
0 0.877806938483627\\
10  0.873294391930149\\
20  0.862259356467587\\
30  0.843519292202805\\
40  0.810975567946546\\
50  0.764952132043205\\
60  0.703789734629508\\
70  0.633709381361381\\
80  0.557880715655734\\
90  0.484447876250541\\
100 0.415453690303134\\
110 0.356031613736596\\
120 0.30477910718066\\
130 0.263947332127982\\
140 0.230198539294046\\
150 0.204749371078216\\
160 0.18389527816686\\
170 0.168900143607271\\
180 0.15632008001432\\
190 0.147760136575114\\
200 0.140121525785821\\
210 0.135320570058077\\
220 0.130521741752366\\
230 0.127855229873028\\
240 0.124654666797007\\
250 0.123181604293945\\
260 0.120870375353601\\
270 0.120059017620999\\
280 0.118239572031144\\
290 0.117793405909043\\
300 0.116244979031429\\
310 0.11599985126903\\
320 0.114600501625605\\
330 0.114465892278431\\
340 0.113148460227259\\
350 0.113074560828397\\
360 0.111802116935638\\
370 0.111761552787695\\
380 0.110513808210221\\
390 0.110491543935363\\
400 0.109257357428493\\
410 0.109245137873159\\
420 0.108018393036357\\
430 0.108011686602734\\
440 0.10678902609503\\
450 0.106785345466141\\
460 0.105564926562271\\
470 0.105562906572213\\
480 0.104343717909652\\
490 0.104342609310153\\
500 0.103124095825278\\
};

% \addplot [color=black]
%   table[row sep=crcr]{%
%   -1 0 \\
%   };
%   \addlegendentry{$\text{AoI}=100$ ms}

%   \addplot [color=black, dashed]
%   table[row sep=crcr]{%
%   -1 0 \\
%   };
%   \addlegendentry{$\text{AoI}=1000$ ms}

%     \addplot [only marks, mark=o,mark size=1.5pt, mark repeat = 3, mark options={solid, green!40!black}]
%   table[row sep=crcr]{%
%   -1 0 \\
%   };
%   \addlegendentry{Safety}

%       \addplot [only marks, mark=triangle,mark size=2.5pt, mark repeat = 3, mark options={solid, red}]
%   table[row sep=crcr]{%
%   -1 0 \\
%   };
%   \addlegendentry{Traffic Management}

%\draw[{Latex[length=1.5mm,width=1.5mm]}-{Latex[length=1.5mm,width=1.5mm]}] (400,0.26)--(400,0.67);
%\node[align=right, rotate=90] at (380,0.47) {\scriptsize VoI drops};
  \node[coordinate] (B) at (axis cs:52,0.75) {};                       % for ellipse
\node[draw, coordinate, align=center, pin={[align=left,pin distance = 5mm, font=\scriptsize\linespread{0.8}\selectfont]-0:{ Safety app \\  more valuable}}] at (axis cs:110,0.75){};  % for pin

\end{axis}
\draw[black, rotate=20] (B) ellipse (0.3 and 0.7);                                % draw ellipse
\end{tikzpicture}%
 \vspace{-0.1cm}
    \caption{\footnotesize Dynamic observations.}
    \label{fig:voi_aoi_dynamic}
      \end{subfigure}
\caption{VoI vs. TX-RX distance and AoI, for different types of observation. Processed VoI operations, $r_h=1080$ px, urban scenarios are considered.}
\label{fig:voi_aoi}
\end{figure}

The following results are derived considering processed VoI assessment operations. 
In Fig.~\ref{fig:voi_observation} we investigate how  VoI evolves as a function of the type of observation.
We observe that  dynamic information is likely to have value for safety applications (although  VoI eventually drops to zero at large distance): the gap between static and dynamic, i.e., short-lived, observations is less than 10\% for $d<100$ m.
Conversely, the large impact of the timeliness attribute in traffic management operations (the weight is 0.6554) makes  VoI decrease as much as 65\% considering dynamic (as opposed to static) observations. 
This makes  sense as time-varying perception records might become obsolete by the time they are actually transmitted.

In Fig.~\ref{fig:voi_aoi} we plot  VoI as a function of the AoI and the sensor reading's temporal characteristics.
Ideally, we would like information to be received as timely as possible, i.e., at the  same instant it was generated by the source. 
However, real-world constraints,  first and foremost the restricted network capacity, put a limit on the frequency at which status updates can be broadcast, thereby making the AoI larger than the inter-transmission time.
For safety services, both ``new'' (i.e., AoI $=100$ ms) and ``old'' (i.e., AoI $=1000$ ms) information provide comparable value to the receiver(s), as long as short-range communications are considered. 
The reason is that  even old  perception records may still potentially increase the opportunity for vehicles to make object detections,  a critical pre-requisite for safety-related operations.
For traffic management services, instead, the impact of the AoI is very strong. 
``Old'' information (i.e., AoI $=1000$ ms) may, in fact, decrease  VoI by more than 70\% considering  dynamic observations  (Fig.~\ref{fig:voi_aoi_dynamic}). To avoid the decay of information, such systems require very frequent  updates to be disseminated through inter-vehicular communications, possibly causing, however, channel access problems.
%although this conflicts with the specific characteristics of the vehicular scenario in which the channel access cannot be centrally managed and transmissions are affected by packet collision.
We notice that, for the dynamic case,  VoI considering traffic management applications is below 0.6 even at short distances: at  $d=10$ m, VoI $=0.58$ vs.  VoI $=0.92$ for safety applications, thereby validating the results in Fig.~\ref{fig:voi_observation}.

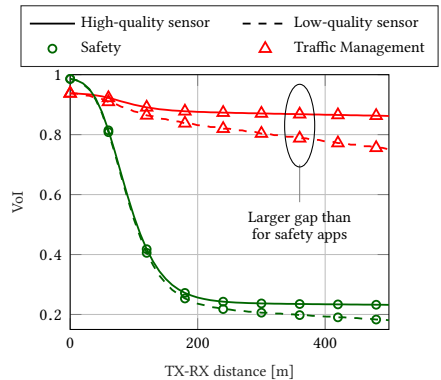
\begin{figure}[t!]
     \centering
     \setlength{\belowcaptionskip}{-0.63cm}
  \setlength\fwidth{0.5\columnwidth}
  \setlength\fheight{0.4\columnwidth}
\pgfplotsset{
tick label style={font=\scriptsize},
label style={font=\scriptsize},
legend  style={font=\scriptsize}
}
\begin{tikzpicture}

\begin{axis}[%
every axis plot/.append style={ line width=0.7pt},
width=\fwidth,
height=\fheight,
at={(0\fwidth,0\fheight)},
scale only axis,
xmin=0,
xmax=500,
xlabel={TX-RX distance [m]},
ymin=0.15,
ymax=1,
ylabel={VoI},
ylabel style={font=\scriptsize\color{white!15!black}},
xlabel style={font=\scriptsize\color{white!15!black}},
axis background/.style={fill=white},
legend columns = {2},
xmajorgrids,
ymajorgrids,
yminorgrids,
legend style={at={(0.5,1.03)}, anchor=south, legend cell align=left, align=left, draw=white!15!black,/tikz/every even column/.append style={column sep=0.35cm}}
]
\addplot [color=red, mark=triangle,mark size=2.5pt, mark repeat = 6, mark options={solid, red},forget plot]
  table[row sep=crcr]{%
0 0.93742817831901\\
10  0.93709655437246\\
20  0.935549559618005\\
30  0.934172365167721\\
40  0.931044711648118\\
50  0.927662480957043\\
60  0.922431660805244\\
70  0.917281503781708\\
80  0.910972869647381\\
90  0.905576340720452\\
100 0.899769967808959\\
110 0.895403080219061\\
120 0.890900531139303\\
130 0.887899832037995\\
140 0.884683618856379\\
150 0.882813376965211\\
160 0.880544786423547\\
170 0.879442804284909\\
180 0.87778226647443\\
190 0.877153202110901\\
200 0.875855808272683\\
210 0.875502989334024\\
220 0.87441428915988\\
230 0.874218329033492\\
240 0.873247084273232\\
250 0.873138829921727\\
260 0.872232941716214\\
270 0.872173315593766\\
280 0.871303568208875\\
290 0.871270779767033\\
300 0.870420949361142\\
310 0.870402935090261\\
320 0.86956406026791\\
330 0.869554167919449\\
340 0.868721313164683\\
350 0.868715882348622\\
360 0.867886333742716\\
370 0.867883352711347\\
380 0.867055619241155\\
390 0.867053983054889\\
400 0.866227245957992\\
410 0.866226347951264\\
420 0.865400157737589\\
430 0.865399664886389\\
440 0.864573774827254\\
450 0.864573504340365\\
460 0.863747779015156\\
470 0.863747630567475\\
480 0.862921995651854\\
490 0.862921914181639\\
500 0.862096328884363\\
};

\addplot [color=green!40!black, mark=o,mark size=1.5pt, mark repeat = 6, mark options={solid, green!40!black},forget plot]
  table[row sep=crcr]{%
0 0.985828731646247\\
10  0.981316185092769\\
20  0.971118313026571\\
30  0.952378248761789\\
40  0.920671687901894\\
50  0.874648251998553\\
60  0.81432301798122\\
70  0.744242664713094\\
80  0.669251162403811\\
90  0.595818322998618\\
100 0.527661300447575\\
110 0.468239223881037\\
120 0.417823880721465\\
130 0.376992105668787\\
140 0.344080476231216\\
150 0.318631308015386\\
160 0.298614378500395\\
170 0.283619243940805\\
180 0.271876343744218\\
190 0.263316400305013\\
200 0.256514952912084\\
210 0.25171399718434\\
220 0.247752332274993\\
230 0.245085820395655\\
240 0.242722420715999\\
250 0.241249358212937\\
260 0.239775292668957\\
270 0.238963934936355\\
280 0.237981652742864\\
290 0.237535486620764\\
300 0.236824223139513\\
310 0.236579095377114\\
320 0.236016909130054\\
330 0.23588229978288\\
340 0.235402031128072\\
350 0.23532813172921\\
360 0.234892851232815\\
370 0.234852287084873\\
380 0.234441705903764\\
390 0.234419441628906\\
400 0.2340224185184\\
410 0.234010198963066\\
420 0.233620617522628\\
430 0.233613911089005\\
440 0.233228413977666\\
450 0.233224733348776\\
460 0.232841477841271\\
470 0.232839457851212\\
480 0.232457432585016\\
490 0.232456323985517\\
500 0.232074973897006\\
};

\addplot [color=red, dashed, mark=triangle,mark size=2.5pt, mark repeat = 6, mark options={solid, red},forget plot]
  table[row sep=crcr]{%
0 0.93742817831901\\
10  0.93709655437246\\
20  0.93109172492817\\
30  0.929714530477886\\
40  0.922129042268448\\
50  0.918746811577372\\
60  0.909058156735738\\
70  0.903907999712202\\
80  0.89314153088804\\
90  0.887745001961111\\
100 0.877480794359783\\
110 0.873113906769885\\
120 0.864153523000291\\
130 0.861152823898984\\
140 0.853478776027533\\
150 0.851608534136365\\
160 0.844882108904865\\
170 0.843780126766227\\
180 0.837661754265913\\
190 0.837032689902384\\
200 0.831277461374331\\
210 0.830924642435672\\
220 0.825378107571693\\
230 0.825182147445305\\
240 0.819753067995209\\
250 0.819644813643704\\
260 0.814281090748356\\
270 0.814221464625908\\
280 0.808893882551182\\
290 0.80886109410934\\
300 0.803553429013614\\
310 0.803535414742733\\
320 0.798238705230547\\
330 0.798228812882086\\
340 0.792938123437485\\
350 0.792932692621424\\
360 0.787645309325682\\
370 0.787642328294313\\
380 0.782356760134286\\
390 0.78235512394802\\
400 0.777070552161288\\
410 0.77706965415456\\
420 0.77178562925105\\
430 0.77178513639985\\
440 0.76650141165088\\
450 0.766501141163991\\
460 0.761217581148947\\
470 0.761217432701265\\
480 0.75593396309581\\
490 0.755933881625594\\
500 0.750650461638483\\
};

\addplot [color=green!40!black, dashed, mark=o,mark size=1.5pt, mark repeat = 6, mark options={solid, green!40!black},forget plot]
  table[row sep=crcr]{%
0 0.985828731646247\\
10  0.981316185092769\\
20  0.969063457417313\\
30  0.950323393152531\\
40  0.916561976683378\\
50  0.870538540780037\\
60  0.808158451153446\\
70  0.73807809788532\\
80  0.661031739966779\\
90  0.587598900561587\\
100 0.517387022401285\\
110 0.457964945834748\\
120 0.405494747065917\\
130 0.364662972013239\\
140 0.329696486966411\\
150 0.30424731875058\\
160 0.282175533626331\\
170 0.267180399066742\\
180 0.253382643260897\\
190 0.244822699821691\\
200 0.235966396819505\\
210 0.23116544109176\\
220 0.225148920573156\\
230 0.222482408693818\\
240 0.218064153404904\\
250 0.216591090901841\\
260 0.213062169748604\\
270 0.212250812016001\\
280 0.209213674213253\\
290 0.208767508091152\\
300 0.206001389000644\\
310 0.205756261238245\\
320 0.203139219381927\\
330 0.203004610034753\\
340 0.200469485770687\\
350 0.200395586371825\\
360 0.197905450266172\\
370 0.19786488611823\\
380 0.195399449327862\\
390 0.195377185053004\\
400 0.192925306333241\\
410 0.192913086777907\\
420 0.19046864972821\\
430 0.190461943294588\\
440 0.18802159057399\\
450 0.188017909945101\\
460 0.185579798828338\\
470 0.185577778838279\\
480 0.183140897962825\\
490 0.183139789363326\\
500 0.180703583665557\\
};

\addplot [color=black]
  table[row sep=crcr]{%
  -1 0 \\
  };
  \addlegendentry{High-quality sensor}

  \addplot [color=black, dashed]
  table[row sep=crcr]{%
  -1 0 \\
  };
  \addlegendentry{Low-quality sensor}

    \addplot [only marks, mark=o,mark size=1.5pt, mark repeat = 3, mark options={solid, green!40!black}]
  table[row sep=crcr]{%
  -1 0 \\
  };
  \addlegendentry{Safety}

      \addplot [only marks, mark=triangle,mark size=2.5pt, mark repeat = 3, mark options={solid, red}]
  table[row sep=crcr]{%
  -1 0 \\
  };
  \addlegendentry{Traffic Management}

  \node[coordinate] (B) at (axis cs:360,0.83) {};                       % for ellipse
\node[draw, coordinate, align=center, pin={[align=center,pin distance = 5mm, font=\scriptsize\linespread{0.8}\selectfont]-90:{ Larger gap than \\ for safety apps}}] at (axis cs:360,0.7){};  % for pin

\end{axis}
\draw[black] (B) ellipse (0.2 and 0.55);                                % draw ellipse
\end{tikzpicture}%
\vspace{-0.33cm}
    \caption{\footnotesize  VoI vs. TX-RX distance for different types of sensors. Processed VoI operations, $P_{\rm td}=1$, AoI 
  ~$=100$~ms, urban scenarios are~considered. }
    \label{fig:voi_sensor}
    \vspace*{-0.1cm}
\end{figure}

The patterns we observed in the previous plots can be recognized also in Fig.~\ref{fig:voi_sensor}, which illustrates  VoI evolution for different types of sensors.
We see that  VoI increases proportionally to the camera resolution, even though the effect of the sensor quality is not very significant:  the gap between high- and low-quality sensor readings is below 15\%.
This is consistent with the outcomes of the AHP comparative evaluations, which  assign very low priority weight to the quality attribute.
In fact, although the network requires context information to be reliable, it still needs to prioritize timely  dissemination to spatially close neighbors to prevent communication failures.
Nevertheless, we observe that sensor quality degradation has a bigger impact on traffic management  than on safety services  (i.e., $w_3=0.2897$ vs. $w_3=0.1336$, respectively).
Finally, Fig.~\ref{fig:voi_sensor} acknowledges the higher \gls{voi} for  safety operations compared to traffic management  at short distance.

%deteriorate

\section{Conclusions and Open Challenges} % (fold)
\label{sec:conclusions_and_open_challenges}

%Frequent and continuous transmissions of sensor information through inter-vehicular communications has recently emerged as a promising opportunity for connected vehicles to  improve  accuracy of detection, recognition and localization of objects and promote advanced safety services. This may, in turn, incur heavy network traffic.
Despite the consensus about using VoI-aware solutions to reduce  link overload,  how to actually realize this is still a largely unexplored issue.
In this paper we investigated for the first time the concept of VoI in vehicular networks and proposed a method that quantifies the expected VoI based on time, space and quality interdependencies.
We evaluated the impact of the operating distance, the type of observation, the type of sensor, the propagation scenario and the age of information on the value assessment.
%We concluded that safety-critical operations should  be prioritized in every working condition, as long as short-range communications are established. 
Moreover, we numerically showed the rate at which  VoI decreases considering obsolete, time-varying and inaccurate~observations.

In this work we demonstrated the feasibility of considering VoI as a proxy for broadcasting decisions in vehicular networks.
As part of our future research, we will validate our framework in dynamic scenarios and investigate the impact of other types of object features (e.g., the road structure) on the end-to-end network performance.
Moreover, we will extend our implementation considering feedback messages from the receiver(s) and learning strategies.

\vspace{-.3cm}
\bibliographystyle{ACM-Reference-Format.bst}
\bibliography{bibliography.bib}

% \balancecolumns
\end{document}